**Title**

# Quantifying EnergyNet performance: a simulation-based framework for decentralized energy networks

**Authors**

*Max Collins[1,2], Jonas Birgersson[2,3], Marc A. Weiss[4,5], Jimmy Chen[6], Mark Z. Jacobson[6], Newsha K. Ajami[7], Daniel Kammen[5,8], Tomas Kåberger[9,10], Franklin Carrero-Martínez[11], Michael Menser[12,13,14], Henrik Ny[9], Lou Riordan[4], Jill G. Ferguson[6]*

**Affiliations**

[1] Lund University

[2] ViaEuropa Sverige AB

[3] Energy Net Task Force

[4] Global Urban Development

[5] University of California, Berkeley

[6] Doerr School of Sustainability, Stanford University

[7] Woods Institute for the Environment, Stanford University

[8] Johns Hopkins University

[9] Blekinge Institute of Technology

[10] Chalmers University of Technology

[11] US National Academies of Sciences, Engineering, and Medicine

[12] Science and Resilience Institute at Jamaica Bay

[13] Brooklyn College

[14] City University of New York

**Corresponding author**

Max Collins

Email: max.collins@iea.lth.se and max.collins@viaeuropa.net

# Quantifying EnergyNet performance: a simulation-based framework for decentralized energy networks

*Max Collins, Jonas Birgersson, Marc A. Weiss, Jimmy Chen, Mark Z. Jacobson, Newsha K. Ajami, Daniel Kammen, Tomas Kåberger, Franklin Carrero-Martínez, Michael Menser, Henrik Ny, Lou Riordan, Jill G. Ferguson*

**Abstract**

*Modern power systems are increasingly exposed to severe disturbances caused by extreme weather, infrastructure failures, and geopolitical conflicts. Conventional power systems rely on interconnected transmission and distribution infrastructure to connect generation and consumers, and failures of this infrastructure can disconnect large groups of consumers from their energy supply until service is restored. At the same time, increasing amounts of local generation, storage, and controllable loads are being deployed throughout the power system, creating new possibilities for how electrical energy systems can be organized and operated.*

*EnergyNet has been proposed as a decentralized, software-defined energy architecture based on interconnected power electronic Energy Routers. Each router coordinates local loads, generation, storage, and grid connections, while bidirectional Energy Links enable energy exchange between routers. Energy exchange is coordinated through the Energy Protocol, allowing routers to communicate energy needs and availability and establish energy transfer across the network. EnergyNet can operate together with the conventional grid, independently from it, or across combinations of grid-connected and isolated regions.*

*This paper develops a quantitative framework for modeling and simulating EnergyNet systems and calculating their performance under different operating conditions and disturbances. The framework represents network topology, prioritized loads, generation, storage, grid connections, Energy Links, and component constraints. Optimization within a time-stepped, receding-horizon simulation calculates feasible energy allocation and exchange. The same framework is also used for disturbance analysis, long-horizon reliability screening, and probabilistic consequence analysis.*

*The framework is first illustrated through a three-router case, demonstrating energy sharing, re-routing, transfer limits, and islanded operation. This case is then extended to demonstrate coordination of a microgrid at the Stanford University campus and an associated off grid microgrid located in Half Moon Bay using an electric bus from the Stanford University bus service as mobile energy resource; under simulated low-solar conditions, scheduled visits support the off grid microgrid while preserving the primary transportation function of the bus.*

*Finally, the framework is applied to a practical, under-construction neighborhood-scale EnergyNet in Lund, Sweden with ten buildings comprising 278 apartments. Full electrical service is maintained during representative summer and winter operation and following the loss of either one grid interface or one Energy Link. Simultaneous loss of two Energy Links isolates part of the network, containing the disturbance while service within the island is determined by local generation, stored energy, load priority, and repair duration. Reliability screening under assumed component failure and repair rates identifies network partitioning as a rare but credible event. Monte Carlo analysis of the network partitioning failure scenario shows full critical service in all sampled summer conditions and a median of approximately 92% continuous critical service under sampled winter conditions.*

## 1. Introduction

The increasing electrification of society, together with the rapid deployment of distributed generation, stationary energy storage, controllable loads, and power electronics, is changing both the structure and operation of modern power systems [1]-[4]. At the same time, extreme weather events, infrastructure failures, geopolitical conflicts, and other large-scale disturbances are placing increasing demands on the ability of electrical infrastructure to maintain essential services [5]-[6].

Conventional power systems rely on interconnected transmission and distribution infrastructure to connect generation and consumers. When critical parts of this infrastructure fail, the consumers supplied through them are disconnected until an alternative supply path is established or the affected infrastructure is restored. Meanwhile, increasing amounts of generation, storage, and controllable demand are being deployed locally throughout the power system [7]-[8]. This creates an opportunity for an energy architecture in which these distributed resources can be coordinated locally and across networks [9]-[10], both together with and independently from the conventional grid.

EnergyNet has been proposed as a decentralized, software-defined energy architecture based on interconnected power electronic **Energy Routers** [11]. It is related to, but distinct from established microgrid, network-microgrid, and Energy Internet architectures. While these fields have established the value of coordinating distributed resources, maintaining islanded operation, and exchanging energy between interconnected local systems [7]-[10], EnergyNet implements these principles through a network of power electronic Energy Routers governed by a common **Energy Protocol** [12]. Each router coordinates local loads, generation, storage, and grid connections, while bidirectional **Energy Links** enable energy exchange between routers. Through the Energy Protocol, routers communicate energy needs and availability and establish energy transfers across the network. EnergyNet can operate together with the conventional grid, independently from it, or across combinations of grid-connected and isolated regions. Local resources can therefore participate in wider energy exchange while retaining local operational capability as system conditions change.

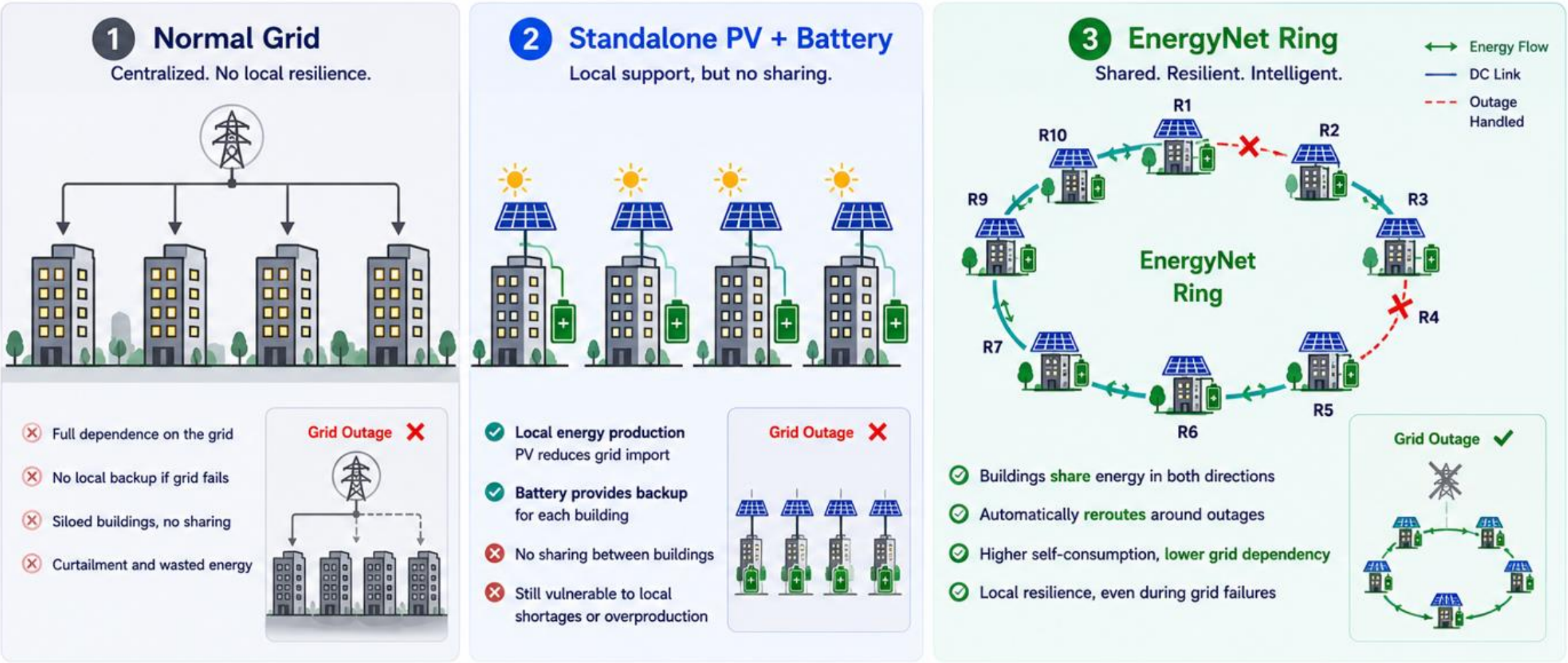

**Figure 1:** Conceptual comparison between conventional electrical distribution and the proposed EnergyNet architecture. EnergyNet enables decentralized coordination of distributed energy resources through interconnected Energy Routers, allowing local energy sharing and adaptive power routing during both normal operation and disturbances.

EnergyNet is currently being explored through laboratory development and commercial pilot installations. However, a general framework for its quantitative analysis is still needed. Its operation involves interactions between distributed generation, storage, loads, grid connections, network topology, and Energy Links, making system performance dependent on both the available resources and the way they are interconnected [9]. Quantitative evaluation therefore requires a model that captures these elements and determines the energy flows that are physically achievable under a given set of conditions.

This paper addresses this need by introducing a unified computational framework for quantitative analysis of EnergyNet systems. An EnergyNet is represented as a network of Energy Routers and Energy Links, with explicit models of loads, generation, storage, grid interfaces, transfer capacities, and component availability. Optimization is used within the simulation to determine feasible energy allocation and exchange subject to these physical constraints and a specified set of priorities. Time-stepped simulation allows EnergyNet performance to be calculated over arbitrary operating periods and conditions, while deterministic disturbances and stochastic component failures extend the same framework to resilience and reliability studies.

The principal contributions of this paper are:

- A general mathematical representation of EnergyNet systems based on interconnected Energy Routers and Energy Links, applicable to arbitrary network topologies.
- A simulation framework for calculating EnergyNet behavior and performance over time, accounting for distributed generation, storage, electrical demand, grid exchange, Energy Link transfers, and physical system constraints.
- A methodology for extending the same framework from normal operation to deterministic disturbances, long-horizon reliability screening, and probabilistic consequence analysis.
- Application of the framework to an under-construction commercial neighborhood-scale EnergyNet with ten interconnected apartment buildings comprising 278 apartments, quantifying its performance under normal operation, representative disturbances, and uncertain operating conditions.

The remainder of the paper is organized as follows. Section 2 introduces the operational abstraction of the EnergyNet architecture and develops the underlying graph representation. Section 3 formulates the optimization-based dispatch problem, including the operational decision variables, physical constraints, and receding-horizon dispatch optimization. Section 4 presents the complete computational framework and simulation methodology. Section 5 demonstrates use of the framework using a series of pedagogical three-router examples that illustrate the principal operational mechanisms of EnergyNet architecture. Section 6 applies the framework to an under-construction commercial neighborhood-scale EnergyNet, evaluating normal operation, deterministic disturbance scenarios, long-horizon reliability screening, and probabilistic consequence analysis. Finally, Section 7 summarizes the principal findings and discusses directions for future research.

## 2. Abstraction and modeling of EnergyNet systems

Quantitative evaluation of an EnergyNet requires a representation that captures the resources, constraints, and energy-transfer paths that determine its system-level performance. The framework focuses on the system-level allocation and exchange of energy rather than the detailed electrical dynamics through which these functions are implemented. Converter switching, voltage regulation, protection, communication protocols, and local controller implementation are therefore not modeled explicitly. Detailed treatment of these topics has been treated extensively in microgrid and power converter control literature, e.g., [13]-[14].

At this level, an EnergyNet is represented as a set of local energy domains interconnected through controllable energy-transfer links. Each domain may contain electrical loads, distributed generation, energy storage, and connections to neighboring domains and the public electricity grid. These domains are represented as Energy Routers and their interconnections as Energy Links. The complete EnergyNet can consequently be represented as a graph $\mathcal{G} = (\mathcal{N}, \mathcal{E})$, in which Energy Routers form the vertices and Energy Links form the edges. This graph representation provides the basis for modeling arbitrary EnergyNet configurations and calculating their performance under different operating conditions.

Figure 2 illustrates the progression from a physical EnergyNet installation to its corresponding Energy Router and graph representations.

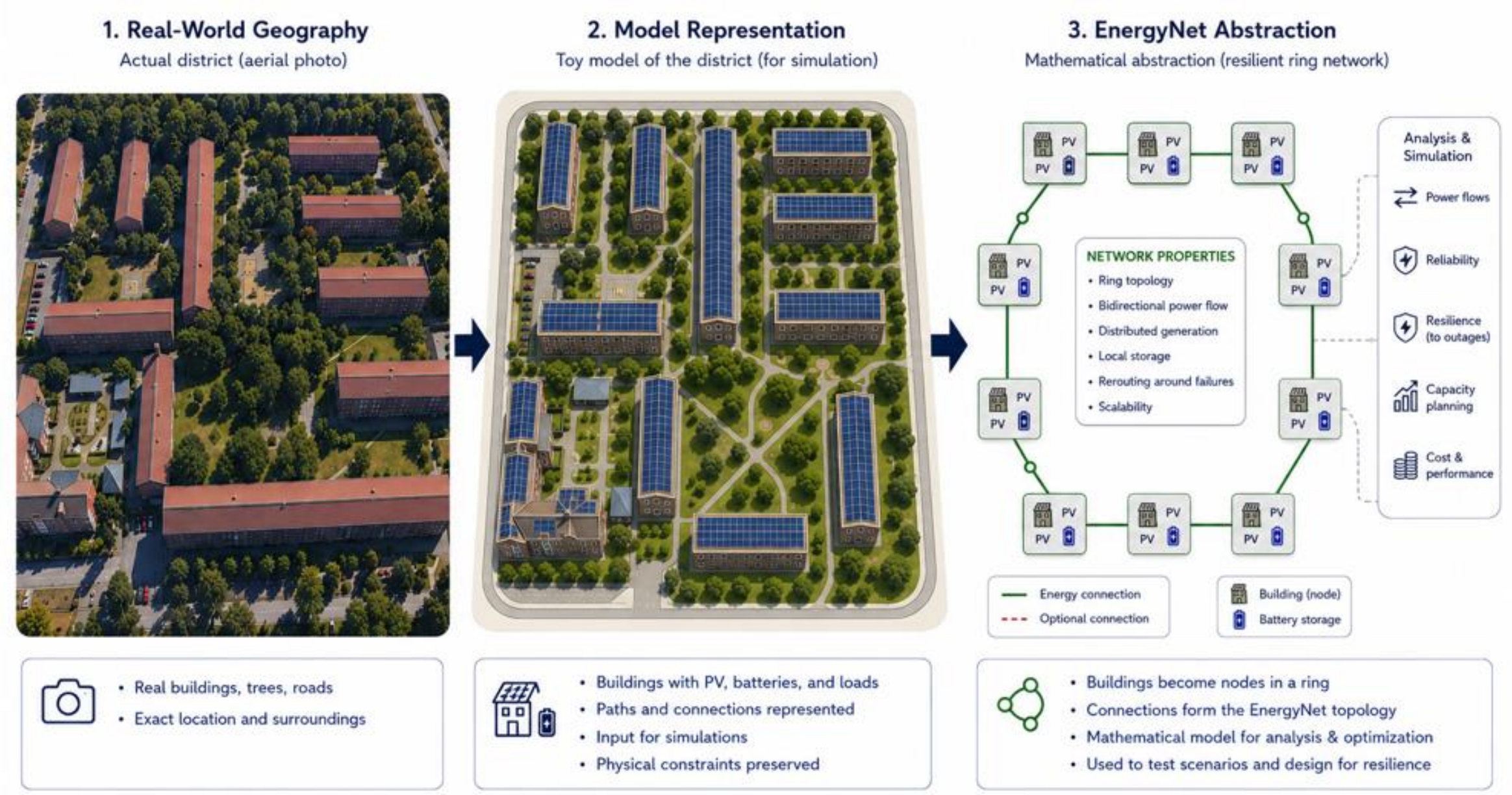

**Figure 2:** Operational abstraction of an EnergyNet system. Physical buildings and local energy resources are represented as Energy Routers interconnected by Energy Links, yielding a graph-based model suitable for optimization, simulation, reliability analysis, and quantitative performance evaluation.

## 2.1 Loads

The primary purpose of an EnergyNet is to supply electrical energy to connected consumers. Within the proposed framework, the electrical demand associated with each Energy Router is represented as a time-varying load profile describing the power requested by the local energy domain at each simulation timestep. A load may represent the aggregated demand of a single building, an apartment block, an industrial facility, or any other autonomous energy domain. Individual appliances and detailed end-user behavior are intentionally not modeled, as the framework focuses on system-level operational decisions rather than detailed consumption dynamics.

Electrical demand at Energy Router $i$ is represented by time-varying load profiles. The framework supports an arbitrary number of priority classes. In the present work, demand is partitioned into **critical** and **flexible** loads denoted $L^c_{i,t}$ and $L^f_{i,t}$, respectively, where critical load represents demand to be maintained whenever physically possible, while flexible load may be reduced or deferred during constrained operation [2], [15].

A distinction is made between **requested** and **served** load. The requested profiles $L^c_{i,t}$ and $L^f_{i,t}$ are external inputs to the simulation, whereas the corresponding served loads $s^c_{i,t}$ and $s^f_{i,t}$ are determined by the dispatch optimization subject to:

$$0 \leq s^c_{i,t} \leq L^c_{i,t} \; ; \; 0 \leq s^f_{i,t} \leq L^f_{i,t}$$

This formulation allows demand to be selectively curtailed when available generation, storage, or network transfer capacity is insufficient, with the relative priority of each load class determined by the operational objective introduced in Section 3.

### 2.2 Energy routers

An Energy Router represents a locally autonomous energy domain that coordinates electrical demand, local generation, energy storage, and, where available, exchange with the public electricity grid. Its internal electrical and control implementation is abstracted, such that the model describes the operational energy flows relevant to system-level analysis.

Figure 3 shows the operational representation adopted in this work. Each router is described by **static parameters**, **time-varying inputs**, **state variables**, and **decision variables**, as summarized in Table 1. Static parameters define the physical operating limits of the router; time-varying inputs describe the imposed operating conditions; state variables couple operation between timesteps; and decision variables determine the resulting dispatch.

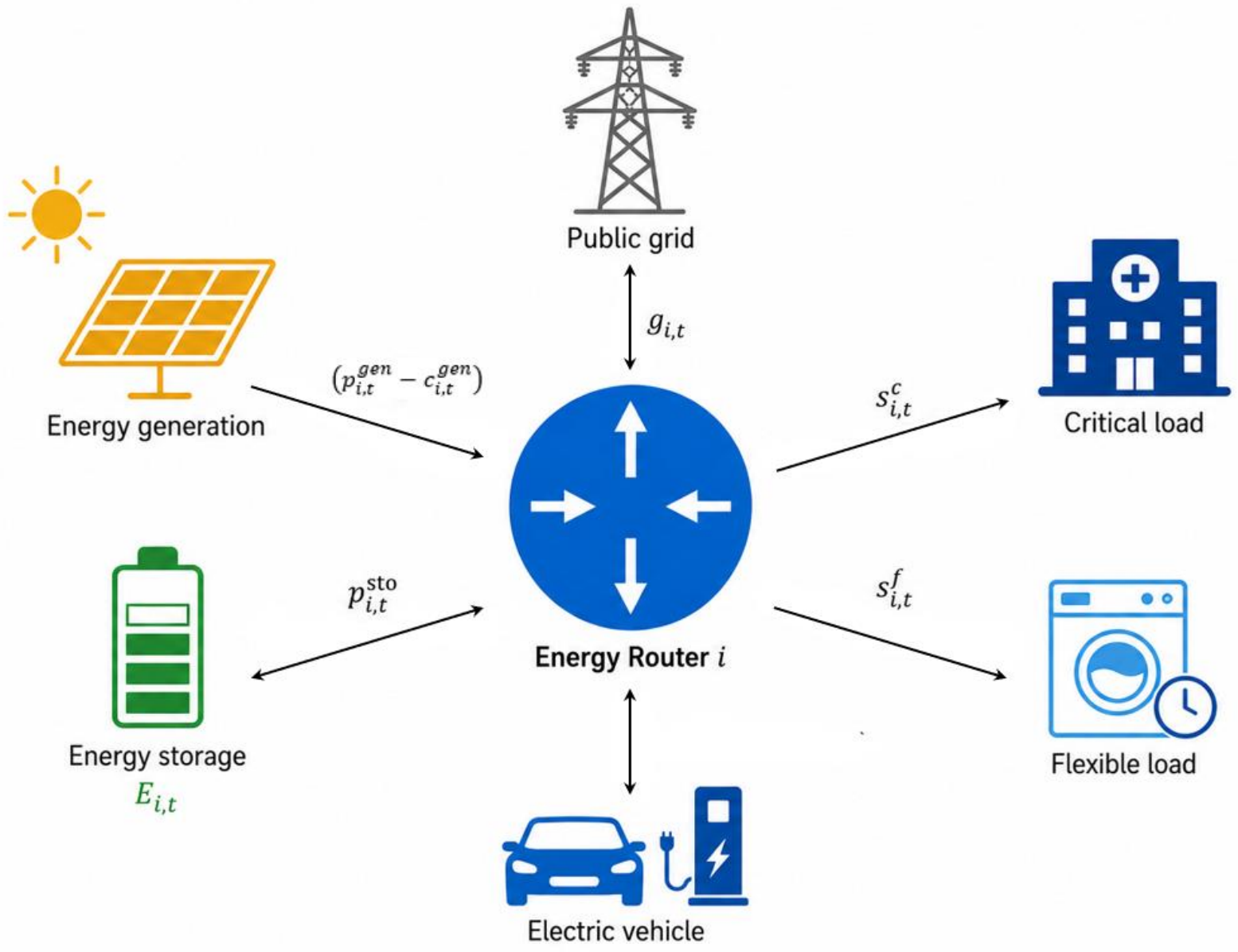

**Figure 3:** Operational representation of an Energy Router. Each router coordinates local energy generation, stationary energy storage, prioritized electrical loads, bidirectional electric-vehicle charging, and an optional public-grid interface.

The operation of an isolated Energy Router is governed by conservation of energy stating that, at every timestep, the total outgoing energy equals the total incoming energy:

$$(p_{i,t}^{gen} - c_{i,t}^{gen}) + p_{i,t}^{sto} + g_{i,t} = s_{i,t}^{c} + s_{i,t}^{f}$$

where positive $g_{i,t}$ denotes grid import and negative values denote export, while positive $p_{i,t}^{\text{sto}}$ denotes storage discharge and negative values denote charging. This local energy balance forms the basis for the interconnected EnergyNet formulation developed in the following section.

**Table 1:** Summary of the parameters, inputs, state variables, and decision variables comprising the proposed Energy Router model.

| Category | Symbol | Unit | Description |
|---|---|---|---|
| Parameter | $E_i^{\max}$ | kWh | Maximum stored energy capacity |
| Parameter | $E_i^{\min}$ | kWh | Minimum permissible stored energy/operational reserve |
| Parameter | $P_i^{ch,\max}$ | kW | Maximum storage charging power |
| Parameter | $P_i^{dis,\max}$ | kW | Maximum storage discharge power |
| Parameter | $\eta$ | - | Storage charging efficiency |
| Parameter | $\eta$ | - | Storage discharge efficiency |
| Parameter | $G_i^{imp,\max}$ | kW | Maximum public grid import power |
| Parameter | $G_i^{exp,\max}$ | kW | Maximum public grid export power |
| Input | $L_{i,t}^{c}$ | kW | Requested critical load power |
| Input | $L_{i,t}^{f}$ | kW | Requested flexible load power |
| Input | $p_{i,t}^{gen}$ | kW | Available local generation power |
| State | $E_{i,t}$ | kWh | Stored energy available |
| Decision | $s_{i,t}^{c}$ | kW | Critical load power served |
| Decision | $s_{i,t}^{f}$ | kW | Flexible load power served |
| Decision | $p_{i,t}^{sto}$ | kW | Storage charge/discharge power |
| Decision | $c_{i,t}^{gen}$ | kW | Curtailed local generation power |
| Decision | $g_{i,t}$ | kW | Public grid import/export power |
| Decision | $q_{i,e,t}$ | kW | Signed Energy Link power exchange |

### 2.3 Energy links

Energy Routers exchange power through controllable Energy Links. Each Energy Link $e \in \mathcal{E}$ is characterized by a maximum transfer capacity $Q_e^{\max}$, conversion efficiency $\eta_e$, and time-dependent availability $A_{e,t}^{\text{link}}$. The power exchanged with router $i$ through an incident Energy Link $e$ is represented by the signed variable $q_{i,e,t}$, where positive values denote power entering the router and negative values denote power leaving it.

An interconnected EnergyNet is represented as a graph $\mathcal{G} = (\mathcal{N}, \mathcal{E})$, where each vertex $i \in \mathcal{N}$ represents an Energy Router and each edge $(i,j) \in \mathcal{E}$ represents an Energy Link. This representation supports arbitrary network topologies independently of their physical scale or geographical layout.

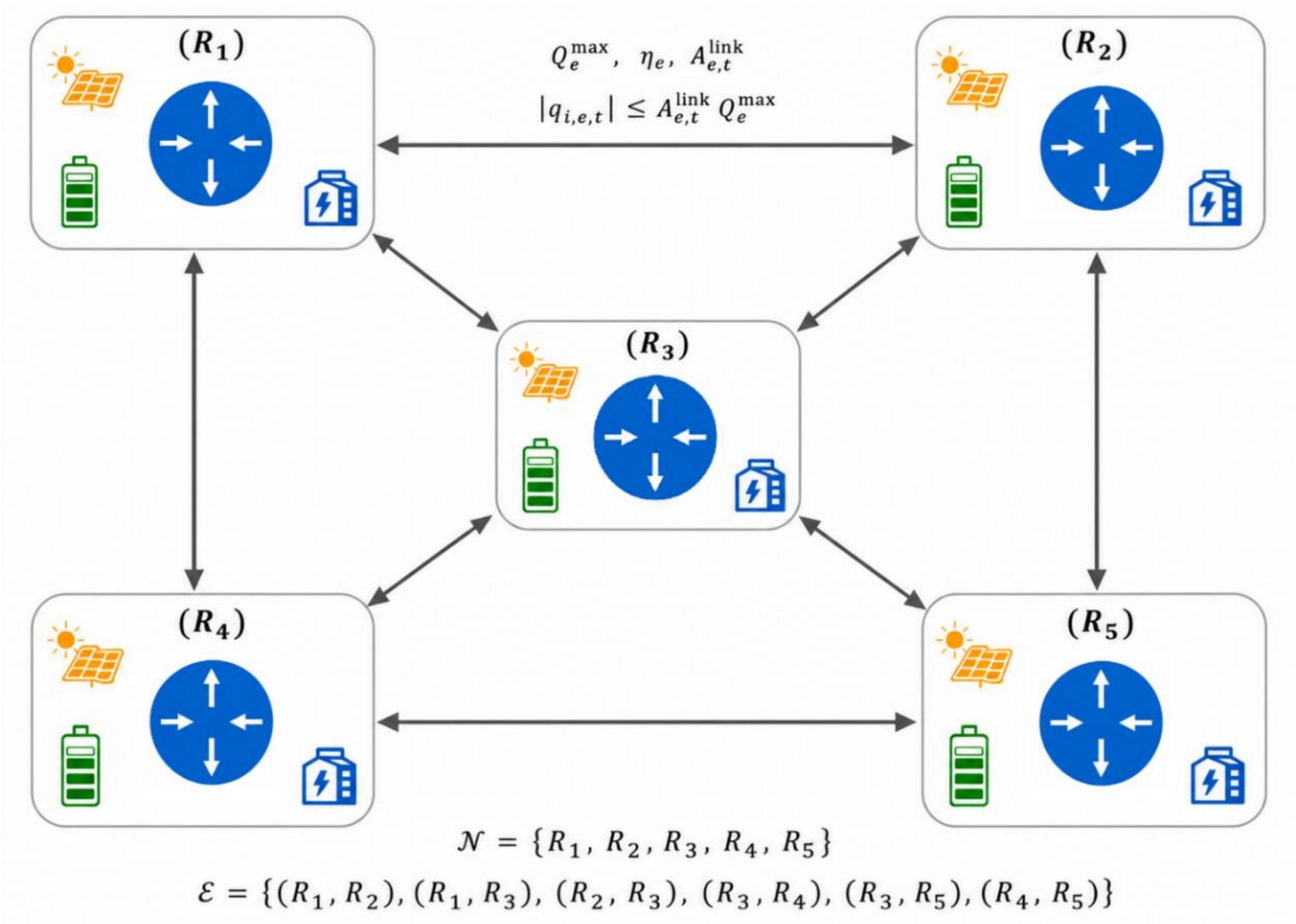

**Figure 4:** Graph representation of an EnergyNet. Energy Routers form the vertex set $\mathcal{N}$ and Energy Links the edge set $\mathcal{E}$. Each Energy Link $e$ is characterized by transfer capacity $Q_e^{\max}$, efficiency $\eta_e$, and availability $A_{e,t}^{\text{link}}$, with $q_{i,e,t}$ denoting the signed power delivered to router $i$ through the link.

Introducing Energy Links extends the local router balance to include energy exchange with neighboring routers:

$$\left(p_{i,t}^{\text{gen}} - c_{i,t}^{\text{gen}}\right) + p_{i,t}^{\text{sto}} + g_{i,t} + \sum_{e \in \delta(i)} q_{i,e,t} = s_{i,t}^{c} + s_{i,t}^{f}$$

where $q_{i,e,t}$ is the signed power delivered to router $i$ through incident Energy Link $e$, positive when entering the router and negative when leaving it. This nodal energy balance, together with the component limits and availability constraints, defines the feasible operation of the EnergyNet and forms the basis for the dispatch formulation in Section 3.

## 3. Optimization-based dispatch

The model developed in Section 2 defines the resources, interconnections, and physical constraints of an EnergyNet. Determining its operation at a given point in time requires deciding how available energy is allocated among electrical loads, storage, grid exchange, and Energy Links. This allocation depends on the current system state, available resources, network constraints, operational priorities, and, where available, information about future conditions.

In a deployed EnergyNet, these decisions are made autonomously by individual Energy Routers coordinating energy exchange through the Energy Protocol. The present framework does not aim to reproduce the details of this distributed decision process. Instead, the physical constraints of the EnergyNet are combined with explicit operational priorities that define the preferred allocation of available energy and resources. The dispatch can therefore be formulated as an optimization problem that selects among feasible energy allocations according to these priorities [16]-[17]. The following sections develop the optimization-based dispatch formulation used to determine this energy allocation as system conditions evolve over time.

### 3.1 Optimization-based dispatch

At each optimization step, the dispatch problem determines the allocation of energy among the decision variables introduced in Section 2, including served load, battery charging and discharging, grid import and export, photovoltaic curtailment, and inter-router energy transfer. Together, these variables define the EnergyNet dispatch for the conditions being considered. The physical constraints define the feasible energy allocations, while the objective function selects among these according to the specified operational priorities.

The objective function can be adapted to the purpose of the study without changing the underlying EnergyNet model. Depending on the application, the dispatch may prioritize objectives such as load service, reduced grid dependence, peak-power reduction, local use of renewable generation, reduced battery utilization, or reduced network transfer. These priorities may also account for anticipated future conditions, allowing present decisions—for example, charging or preserving stored energy—to reflect expected changes in demand, generation, or system availability. The treatment of such forecast information is introduced in Section 3.3.

In the present work, the objective prioritizes electrical load service, with critical demand assigned higher priority than flexible demand. Additional terms govern the preferred use of grid exchange, battery storage, Energy Link transfers, photovoltaic curtailment, and peak grid import. The objective is formulated as:

$$\max J$$

where:

$$J = \sum_t \left[ w_c \sum_i s_{i,t}^c + w_f \sum_i s_{i,t}^f - \alpha_{\text{imp}} \sum_i [g_{i,t}]^+ + \alpha_{\text{exp}} \sum_i [g_{i,t}]^- - \alpha_{\text{bat}} \sum_i |p_{i,t}^{\text{sto}}| - \alpha_{\text{flow}} \sum_{e \in \mathcal{E}} |q_{e,t}| - \alpha_{\text{curt}} \sum_i c_{i,t}^{\text{pv}} - \alpha_{\text{peak}} P_{\text{grid}}^{\text{peak}} \right]$$

The weights $w_c$ and $w_f$ define the relative priority of the load classes, while the remaining coefficients penalize or reward grid exchange, battery utilization, EnergyNet transfer, photovoltaic curtailment, and peak grid import. These coefficients represent operational preferences rather than economic costs and determine how available flexibility is utilized when multiple feasible dispatch solutions exist.

### 3.2 Constraints

The EnergyNet model developed in Section 2 defines the physical resources and interconnections available to the dispatch. These are expressed as constraints that define the feasible energy allocations from which the objective function selects the preferred dispatch. For each router $i$, Energy Link $(i, j)$, and timestep $t$, the constraints are defined as follows:

**Energy balance.** Conservation of energy at each Energy Router requires for all $i$:

$$\left(p_{i,t}^{\text{gen}} - c_{i,t}^{\text{gen}}\right) + p_{i,t}^{\text{sto}} + g_{i,t} + \sum_{e \in \delta(i)} q_{i,e,t} = s_{i,t}^{c} + s_{i,t}^{f}$$

**Load service.** Served demand is bounded by the corresponding requested demand:

$$0 \leq s_{i,t}^{c} \leq L_{i,t}^{c}; 0 \leq s_{i,t}^{f} \leq L_{i,t}^{f}$$

**Battery operation.** Charging and discharging powers are constrained by:

$$-P_i^{\text{ch,max}} \leq p_{i,t}^{\text{sto}} \leq P_i^{\text{dis,max}}$$

while stored energy evolves according to:

$$E_{i,t+1} = E_{i,t} + \eta_i^{\text{ch}} [p_{i,t}^{\text{sto}}]^{-} \Delta t - \frac{\left[p_{i,t}^{\text{sto}}\right]^{+}}{\eta_i^{\text{dis}}} \Delta t$$

$$E_i^{\text{min}} \leq E_{i,t} \leq E_i^{\text{max}}.$$

**Grid interface.** Grid exchange is constrained by the interface rating and its availability:

$$-A_{i,t}^{\text{grid}} G_i^{\text{exp,max}} \leq g_{i,t} \leq A_{i,t}^{\text{grid}} G_i^{\text{imp,max}}$$

**Energy Links.** Inter-router power transfer is constrained by link capacity and availability:

$$-A_{e,t}^{\text{link}} Q_e^{\max} \leq q_{e,t} \leq A_{e,t}^{\text{link}} Q_e^{\max}$$

**Peak grid import.** When peak import is included, the auxiliary variable $P_{\text{grid}}^{peak}$ must satisfy:

$$P_{\text{grid}}^{\text{peak}} \geq \sum_i [\, g_{i,t}]^+, \forall t$$

Together, these constraints ensure that each calculated dispatch remains consistent with the physical capabilities and availability of the EnergyNet.

### 3.3 Forecasting and receding-horizon dispatch

Decisions involving stored energy and other time-coupled resources may benefit from information about future operating conditions. To support such predictive dispatch, the framework distinguishes between **actual** and **forecast** system variables, [16]-[17]. Actual values describe the realized operating conditions and determine the evolution of the system, whereas forecast values provide estimates of future conditions that may influence present decisions.

A forecast of a generic variable $x$, made at timestep $t$ for a future timestep $t + k$, is denoted by $\hat{x}_{t+k|t}$, where the notation indicates that the estimate of the value at $t + k$ is based on information available at $t$. Forecast quantities may include electrical demand, photovoltaic generation, grid availability, and Energy Link availability. The information available to the optimization at timestep $t$ considering an optimization horizon $H$ can therefore be represented as:

$$\mathcal{I}_t = \{x_t, \hat{x}_{t+1|t}, \hat{x}_{t+2|t}, \dots, \hat{x}_{t+H-1|t}\}$$

This separation between actual and forecast information allows anticipated conditions, such as scheduled outages or expected variations in demand and generation, to influence dispatch before they occur. Unanticipated disturbances, in contrast, affect operation only once they are reflected in the realized system state. Forecast uncertainty is not modeled explicitly within the optimization formulation but may be introduced through direct forecast errors.

Forecast information is incorporated into the optimizer through a receding-horizon dispatch strategy. At each timestep $t$, the optimizer uses the current system state and forecasts over the horizon $H$ to determine a sequence of dispatch decisions:

$$\{u_t, u_{t+1}, \dots, u_{t+H-1}\},$$

where $u_t$ denotes the collection of decision variables defining the EnergyNet dispatch. Although the optimization determines a dispatch over the complete horizon, only the decisions for the current timestep are implemented.

Then, the system state is updated using the realized operating conditions, the horizon advances by one timestep, and the optimization is repeated using the updated state and available forecasts. In this way, present decisions can account for anticipated future conditions while continuously adapting to the actual evolution of the EnergyNet. Figure 5 illustrates this process, in which each optimization produces one implemented dispatch and a provisional plan for the remainder of the horizon.

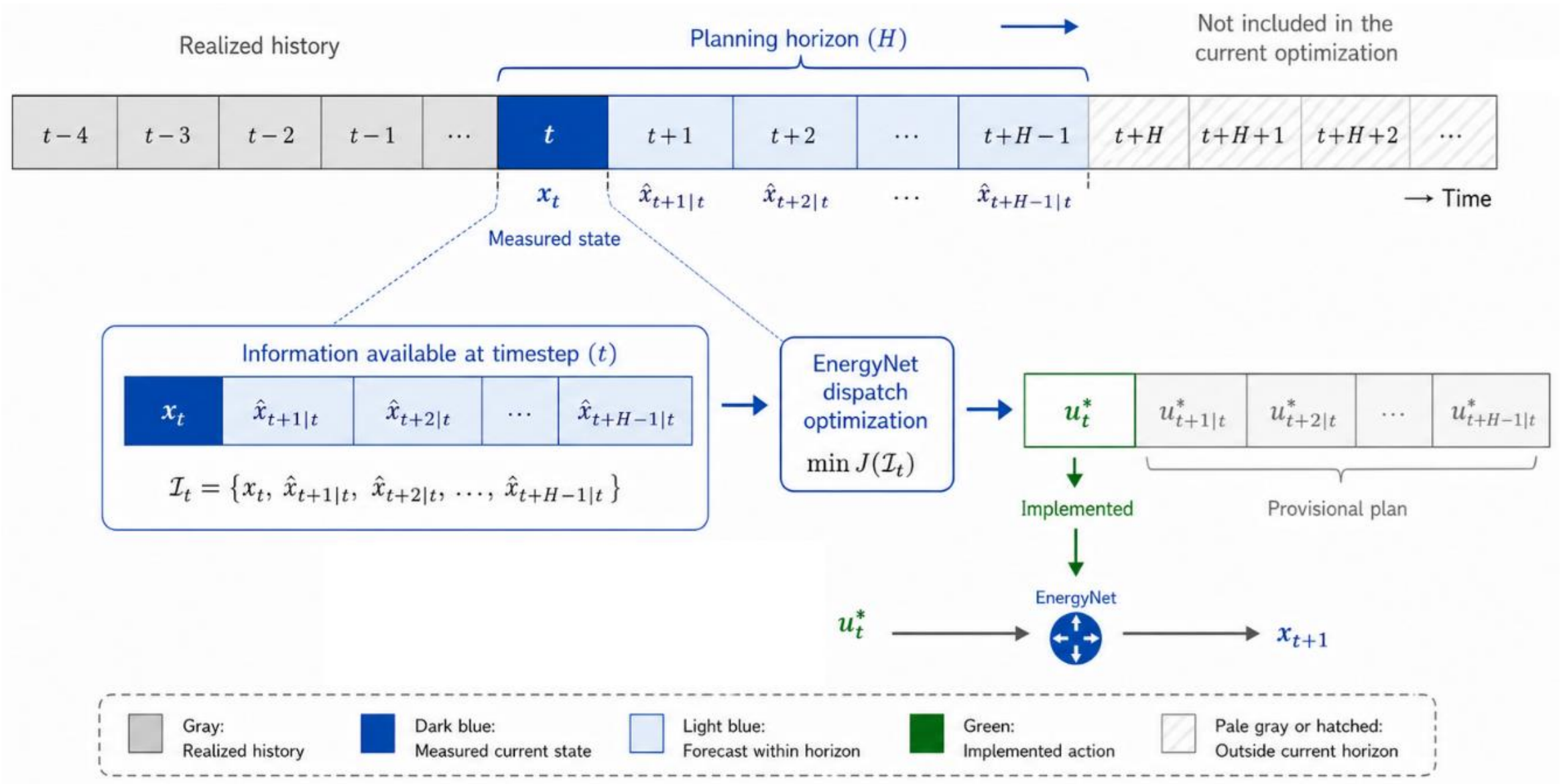

**Figure 5:** Receding-horizon dispatch over a finite planning window. At timestep $t$, the dispatch optimization combines the current EnergyNet state $x_t$ with forecasts extending to $t + H - 1$. The resulting sequence contains one implemented action $u_t^*$ and a provisional plan for the remainder of the horizon. After the first action is applied, the realized state is updated, the planning window advances, and the optimization is repeated with updated information.

The formulation encompasses both myopic and predictive operation. For $H = 1$, dispatch depends only on the current system state, whereas $H > 1$ allows future operating conditions to influence present decisions through the available forecasts.

### 3.4 Emergency reserve and sustainable service commitment

The dispatch formulation described above prioritizes critical over flexible demand when available resources are constrained. During prolonged disturbances, however, the available energy may be insufficient to maintain full critical-load service throughout the disturbance, making the scheduling and preservation of stored energy an important aspect of resilient operation [18]. The framework allows emergency policies to be defined, including how stored energy is protected before a disturbance and how energy is allocated once emergency operation is required.

A specified fraction of energy may be protected during normal operation as emergency reserve:

$$E_{i,t} \geq E_i^{\text{res}}$$

where $E_i^{\text{res}}$ is the protected reserve level at router $i$. This energy is unavailable to normal dispatch but may be released when a defined disturbance is detected. When a disturbance creates an electrically isolated section, the framework can determine the critical-load service that can be sustained over the expected repair interval according to:

$$s_{i,\tau}^{c} = \gamma L_{i,\tau}^{c}, i \in \mathcal{N}^{\text{isl}}, \tau \in \mathcal{T}^{\text{rep}}$$

where $\mathcal{N}^{\text{isl}}$ denotes the routers within the island, $\mathcal{T}^{\text{rep}}$ the expected repair interval, and $\gamma \in [0,1]$ the critical-service fraction. The largest feasible value of $\gamma$ may be determined from the available stored energy, forecast demand and generation, and the physical constraints of the island. Flexible demand is curtailed as required to maintain this commitment; remaining energy may subsequently be allocated according to the specified emergency policy.

The commitment may be recalculated within the receding-horizon framework as the system state, forecasts, connectivity, or expected repair time change. The emergency policy therefore provides a means of preserving and allocating existing resources during prolonged disturbances without introducing additional generation, storage, or network capacity. Figure 6 summarizes the approach.

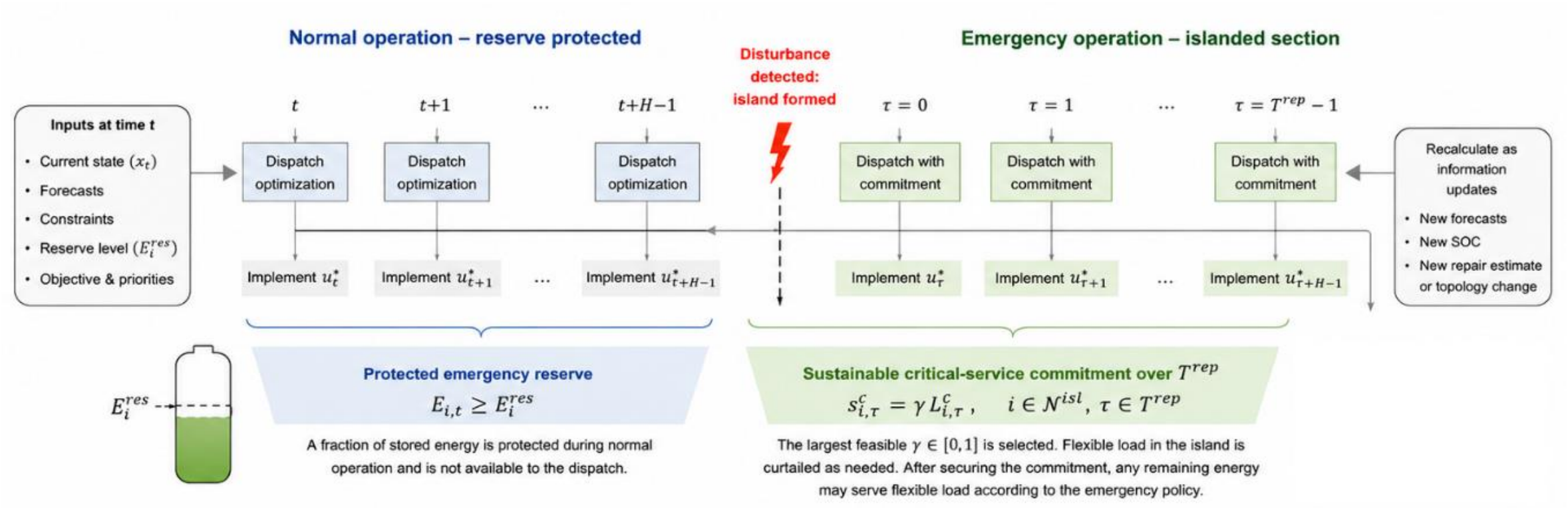

**Figure 6:** Emergency reserve and sustainable service commitment. Stored energy is protected during normal operation and released following islanding to maintain a sustainable fraction of critical load over the expected repair interval.

## 4. Simulation and evaluation framework

The EnergyNet model and dispatch formulation developed in Sections 2 and 3 are combined within a time-stepped simulation framework to evaluate EnergyNet behavior and performance over a specified study period. A study is defined by the EnergyNet configuration, time-series inputs for demand and generation, operational objective and planning horizon, and any disturbance or forecast assumptions under investigation. Together, these inputs define the physical system and operating conditions to which the dispatch formulation is applied.

At each timestep, the current system state and available forecasts are assembled, and the dispatch problem is solved over the applicable planning horizon. Only the first optimized action is implemented, after which the system state is updated according to the realized operating conditions. The implemented dispatch and resulting state are recorded, and the procedure is repeated over the specified simulation period. Figure 7 summarizes the overall framework from study definition through simulation to performance evaluation.

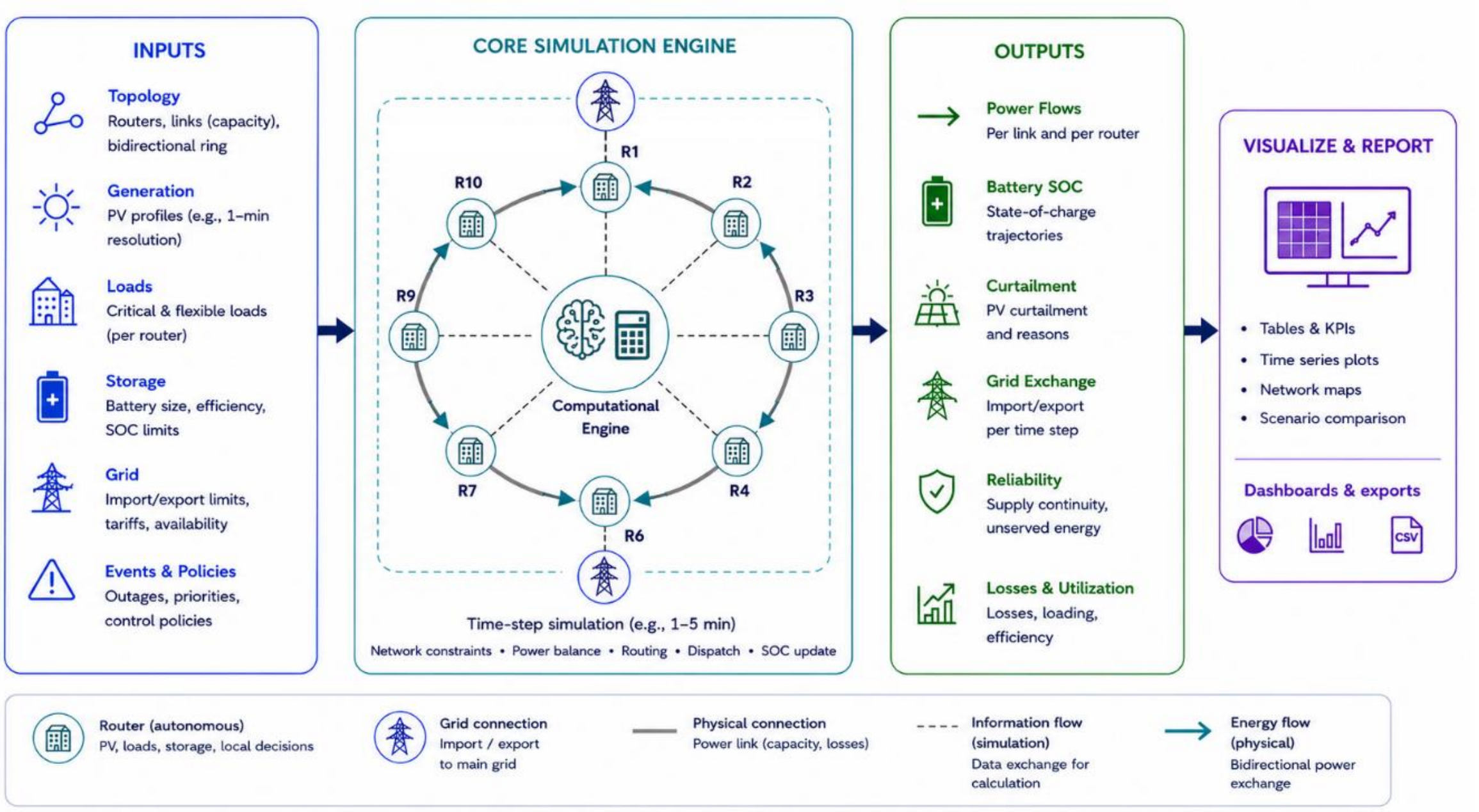

**Figure 7:** General simulation and evaluation framework for EnergyNet studies. The reusable EnergyNet model and optimization-based dispatch framework are combined with a study-specific system instance, operating conditions, and operational assumptions. Time-stepped simulation produces a complete operational record from which application-specific performance metrics are evaluated. Deterministic, parametric, and Monte Carlo studies modify or sample the study definition while preserving the same underlying model and simulation procedure.

The complete simulation procedure is summarized in Algorithm 1. The inputs comprise the EnergyNet graph $\mathcal{G} = (\mathcal{N}, \mathcal{E})$, Energy Router and Energy Link parameters, realized and forecast time series, initial storage states, objective and constraint parameters, planning horizon $H$, and timestep duration $\Delta t$. The simulation returns the realized operational trajectory $Y$ and a set of application-specific performance metrics $M$.

**Algorithm 1:** Pseudocode for the time-stepped receding-horizon EnergyNet simulation

| Step | Description |
|---|---|
| **Input** | EnergyNet graph $\mathcal{G} = (\mathcal{N}, \mathcal{E})$; Energy Router and Energy Link parameters; realized time-series $X$; forecast time-series $\hat{X}$; initial stored energy $E_{i,0}$; objective and constraint parameters; planning horizon $H$; timestep duration $\Delta t$. |
| **Output** | Realized operational trajectory $Y$; application-specific performance metrics $M$. |

| Step | Description |
|---|---|
| **1** | Initialize the stored-energy state $E_{i,0}$ for all $i \in \mathcal{N}$. |
| **2** | Initialize an empty operational trajectory $Y$. |
| **3** | **for** $t = 0, \dots, T-1$ **do** |
| | Compute the effective planning horizon $h = min(H, T-t)$. |
| | Read the realized operating conditions $x_t$. |
| | Assemble the information set $I_t = \{x_t, \hat{x}_{t+1\|t}, \dots \hat{x}_{t+h-1\|t}\}$, where $x_t$ is realized and all subsequent states are forecasts. |
| | Apply realized and forecast component availability by assigning zero transfer capacity to unavailable public-grid interfaces and unavailable Energy Links. |
| | Formulate the dispatch optimization over $\tau = t, \dots, t+h-1$, including served load, storage operation, grid exchange, local-generation curtailment, and Energy Link power flows. |
| | Apply the physical and operational constraints together with the selected operational objective. |
| | Solve the optimization to obtain $U_t^* = \{u_t^*, u_{t+1\|t}^*, \dots, u_{tt+h-1\|t}^*\}$. |
| | Implement only the first optimized action $u_t^*$; the remaining actions remain provisional until the next timestep. |
| | Update the stored-energy state according to $E_{i,t+1} = E_{i,t} + \eta_i^{\mathrm{ch}}[p_{i,t}^{\mathrm{sto}}]^{-}\Delta t - \frac{[p_{i,t}^{\mathrm{sto}}]^{+}}{\eta_i^{\mathrm{dis}}}\Delta t$ |
| | Record the realized operational quantities in $Y$. |
| **4** | **end for** |
| **5** | Compute the application-specific performance metrics (M) from the complete operational trajectory $Y$. |
| **6** | Return $Y$ and $M$. |

The completed trajectory $Y$ provides a time-resolved description of EnergyNet operation, including served demand, storage operation, grid exchange, photovoltaic curtailment, and Energy Link power flows. Application-specific performance metrics $M$ are subsequently derived from this trajectory; in the present work, these include load service, unserved energy, grid exchange, photovoltaic curtailment, battery utilization, Energy Link loading, and peak grid import.

Repeated execution under different operating conditions, disturbance scenarios, or sampled reliability histories extends the same procedure from individual simulations to comparative, reliability, and probabilistic analyses [19].

## 5. Illustrative case: three-router EnergyNet

The framework developed in the preceding sections is first applied to a deliberately simple three-router EnergyNet in which the resulting energy flows can be interpreted directly from the underlying energy balances. The purpose is to illustrate how the framework represents the principal mechanisms of EnergyNet operation before applying it to the larger neighborhood-scale system in Section 6.

The study progresses from normal operation to increasingly constrained conditions involving Energy Link outages, transfer limits, energy storage, islanding, and local generation. The same basic system configuration is retained throughout, with individual parameters and component availability modified between cases to isolate each mechanism.

### 5.1 Common three-router configuration

The illustrative EnergyNet consists of three Energy Routers $\mathcal{N} = \{R_1, R_2, R_3\}$ interconnected by three Energy Links $\mathcal{E} = \{(R_1, R_2), (R_2, R_3), (R_3, R_1)\}$ forming the triangular network shown in Figure 8. Router $R_1$ is connected to the public grid and may import or export power, while $R_2$ and $R_3$ interact with the grid only indirectly through the EnergyNet. Constant critical loads of 2, 4, and 3 kW are assigned to $R_1$, $R_2$, and $R_3$, respectively, giving a total critical demand of 9 kW. Flexible demand is omitted to isolate the routing and energy-balancing mechanisms considered in these examples.

Each case is simulated for 4 hours with a 15 min timestep and a planning horizon of $H = 1$, such that dispatch depends only on the current operating conditions. Energy Link efficiencies are set to unity to isolate the effects of topology and transfer capacity from losses. The objective prioritizes load service while applying a small penalty to unnecessary Energy Link power flow.

This configuration is retained throughout the entire section. Local generation, storage, link capacities, and component availability are modified only as required by each case, allowing changes in the resulting dispatch to be associated directly with the mechanism under investigation.

### 5.2 Normal energy routing

The first case establishes the reference operating condition for the subsequent scenarios. All three Energy Links are available with a transfer capacity of 20 kW, while no local generation or storage is present. Consequently, the public-grid connection at $R_1$ supplies the entire 9 kW critical demand of the network.

The resulting dispatch is shown in Figure 8. Here, router $R_1$ imports 9 kW, supplies its local 2 kW load, and transfers the remaining power directly to $R_2$ and $R_3$, which receive 4 and 3 kW, respectively. No power is transferred between $R_2$ and $R_3$. All load demand is therefore supplied, corresponding to 100% critical-load service and a total grid import of 36 kWh over the 4-hour simulation.

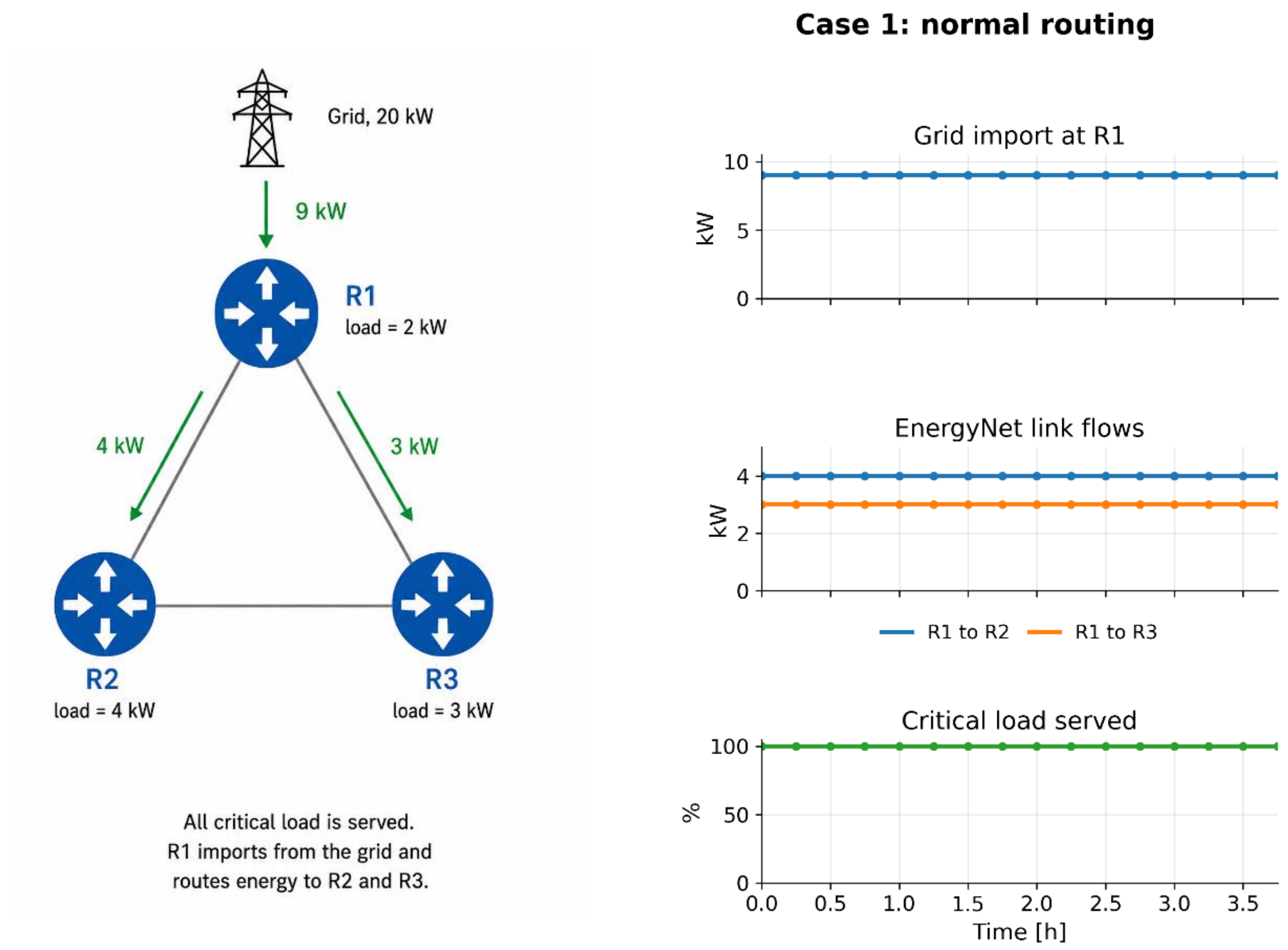

**Figure 8:** Normal network operation with all Energy Links available. The optimization imports power through the public-grid interface at $R_1$ and supplies the critical loads at all three Energy Routers using the direct transmission paths.

Importantly, these transfer paths are not prescribed by the simulation. They emerge from the network-wide dispatch optimization subject to the EnergyNet topology, physical constraints, and operational objective. This baseline therefore establishes how the power-flow results should be interpreted in the subsequent cases, where changes in network conditions cause the dispatch and routing pattern to adapt accordingly.

### 5.3 Automatic rerouting following an Energy Link outage

The second case introduces a failure of the direct Energy Link between $R_1$and $R_2$, while all other conditions remain unchanged. Router $R_2$ therefore loses its direct connection to the grid-connected router but remains connected to $R_3$. The resulting dispatch is shown in Figure 9. The 4 kW required by $R_2$ is automatically rerouted through the surviving path $R_1 \rightarrow R_3 \rightarrow R_2$. Router $R_1$ continues to import the full 9 kW network demand, and all critical loads remain supplied despite the link outage.

This case demonstrates the role of **network redundancy** in EnergyNet operation. No alternative route is prescribed in advance; when the failed link becomes unavailable, the optimization determines a new feasible dispatch over the remaining topology. The outage therefore changes the spatial distribution of power flow without affecting load service, provided sufficient transfer capacity remains along the surviving path.

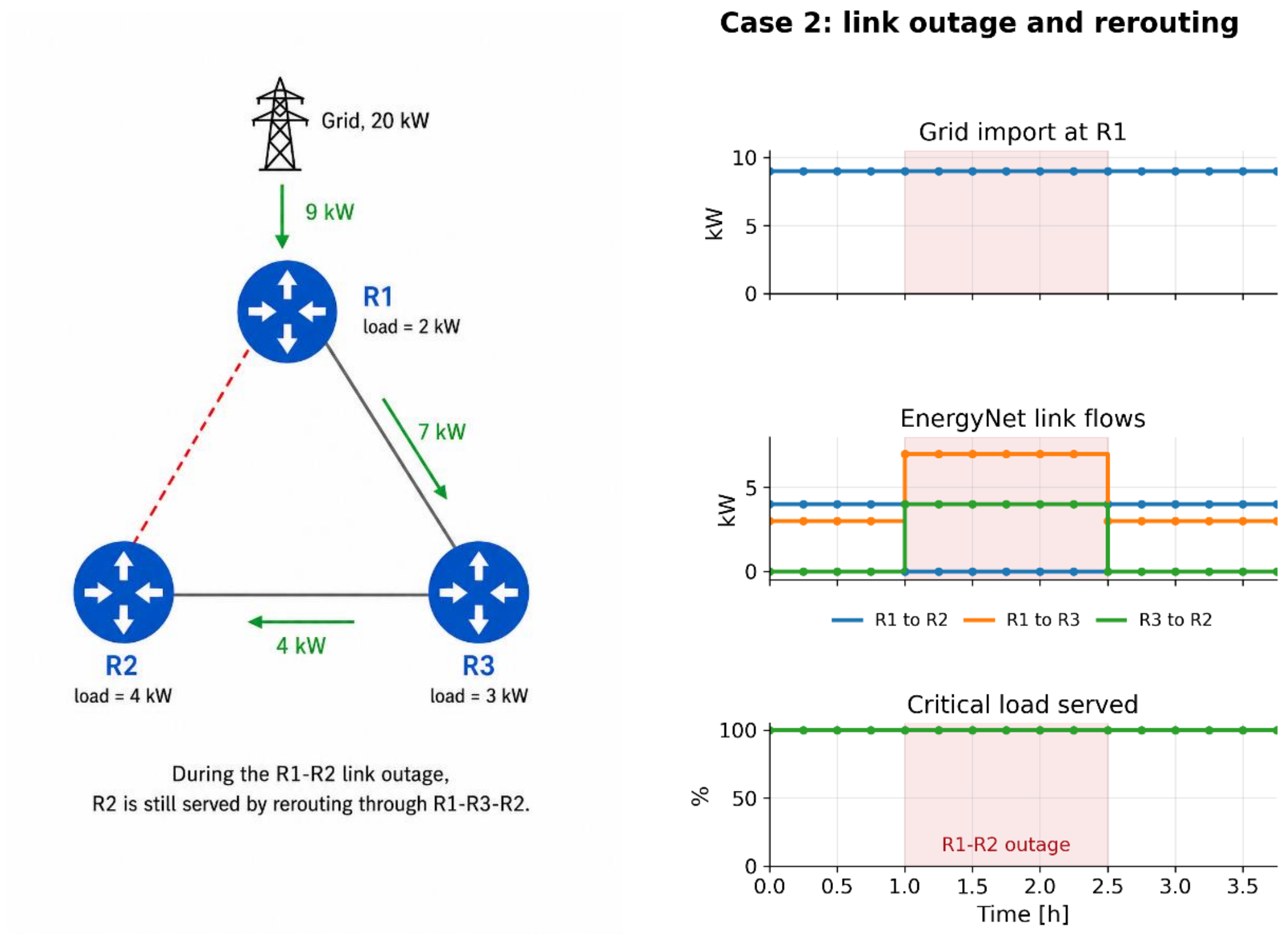

**Figure 9:** Operation following the temporary outage of the $R_1$–$R_2$ Energy Link. The optimization automatically reroutes power through the remaining path $R_1 \rightarrow R_3 \rightarrow R_2$, maintaining uninterrupted critical-load service without predefined contingency-routing logic.

### 5.4 Link outage and capacity-limited rerouting

The third case retains the $R_1 - R_2$ link outage introduced in Section 5.3 but reduces the transfer capacity of the surviving $R_1 - R_3$ link from 20 kW to 5 kW. The alternative route to $R_2$ therefore remains available, but its capacity is insufficient to supply the full combined demand of $R_2$ and $R_3$.

As shown in Figure 10, the $R_1 - R_3$ link operates at its 5 kW capacity. Router $R_3$ supplies its local 3 kW demand and transfers the remaining 2 kW to $R_2$. Since $R_2$ requests 4 kW, only half of its critical demand can be supplied.

This case illustrates **graceful degradation** under constrained operation. Although the surviving network cannot maintain full service, the EnergyNet continues to use the available transfer capacity to supply as much demand as physically possible. The capacity constraint therefore results in partial curtailment rather than complete loss of service at the affected router.

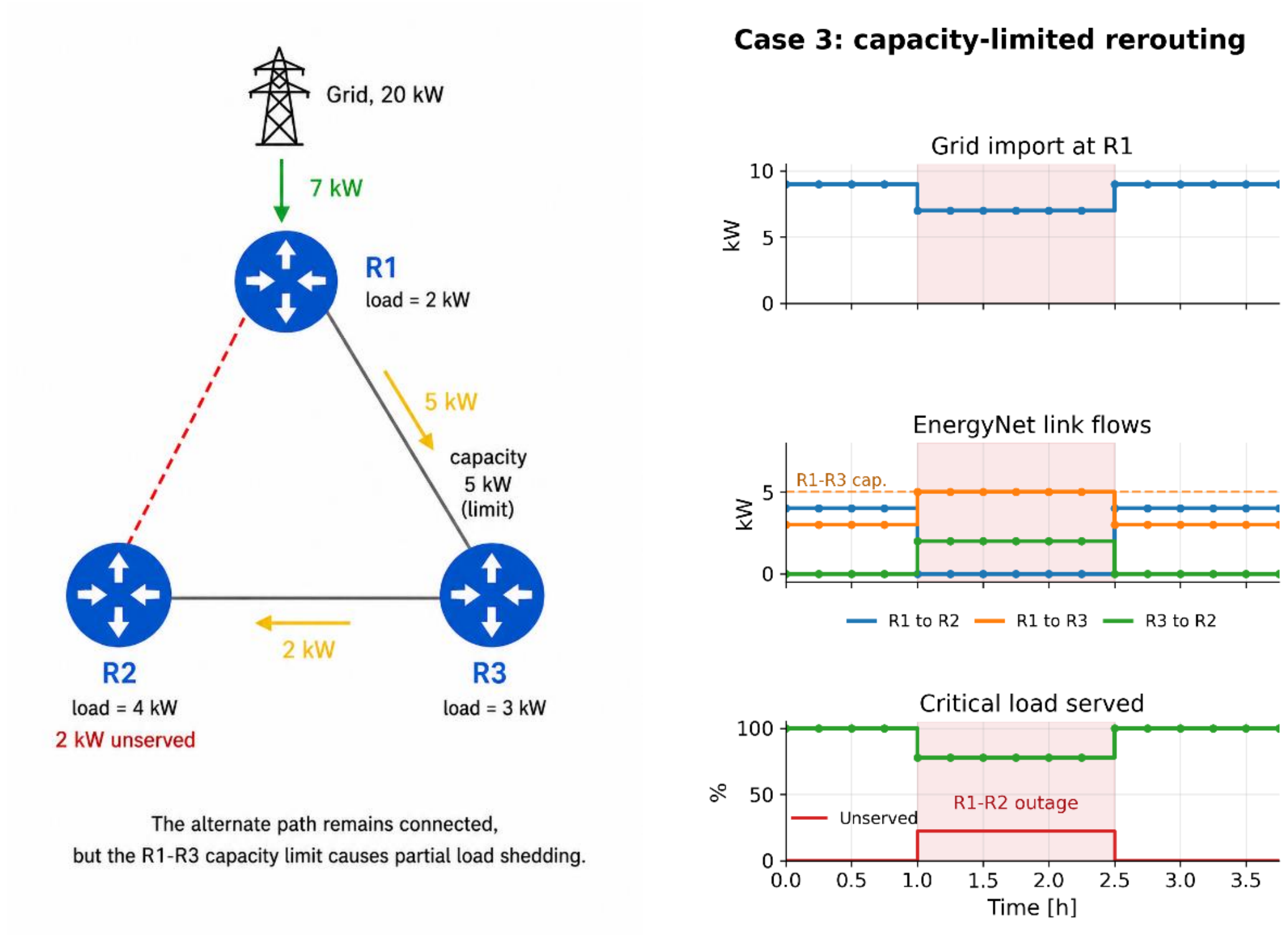

**Figure 10:** Operation with the $R_1$–$R_2$ Energy Link unavailable and the remaining $R_1$–$R_3$ Energy Link limited to 5 kW. Although the network remains connected, the available transfer capacity is insufficient to supply the complete critical load, resulting in partial service at $R_2$.

### 5.5 Storage-supported rerouting and islanding

The fourth case extends the capacity-constrained condition of Section 5.4 by introducing stationary energy storage at $R_2$. The storage system has a capacity and initial stored energy of 8 kWh and a maximum charge and discharge power of 4 kW. The $R_1 - R_3$ link remains limited to 5 kW, while the $R_1 - R_2$ link becomes unavailable at $t = 1.00$ h as in the previous case. At $t = 2.50$ h, the $R_2 - R_3$ link also becomes unavailable, completely islanding $R_2$; both links are restored at $t = 3.50$ h.

During the first disturbance stage, the surviving network can again transfer only 5 kW from $R_1$ to $R_3$. Of this, 3 kW supplies the local load at $R_3$ and 2 kW is transferred onward to $R_2$. Unlike the previous case, however, the remaining 2 kW required by $R_2$ is supplied by its local storage, allowing its full 4 kW critical load to remain served despite the network constraint.

When the second link fails at $t = 2.50$ h, $R_2$ becomes electrically isolated from the rest of the EnergyNet. The battery then supplies the complete 4 kW local critical load until network connectivity is restored at $t = 3.50$ h.

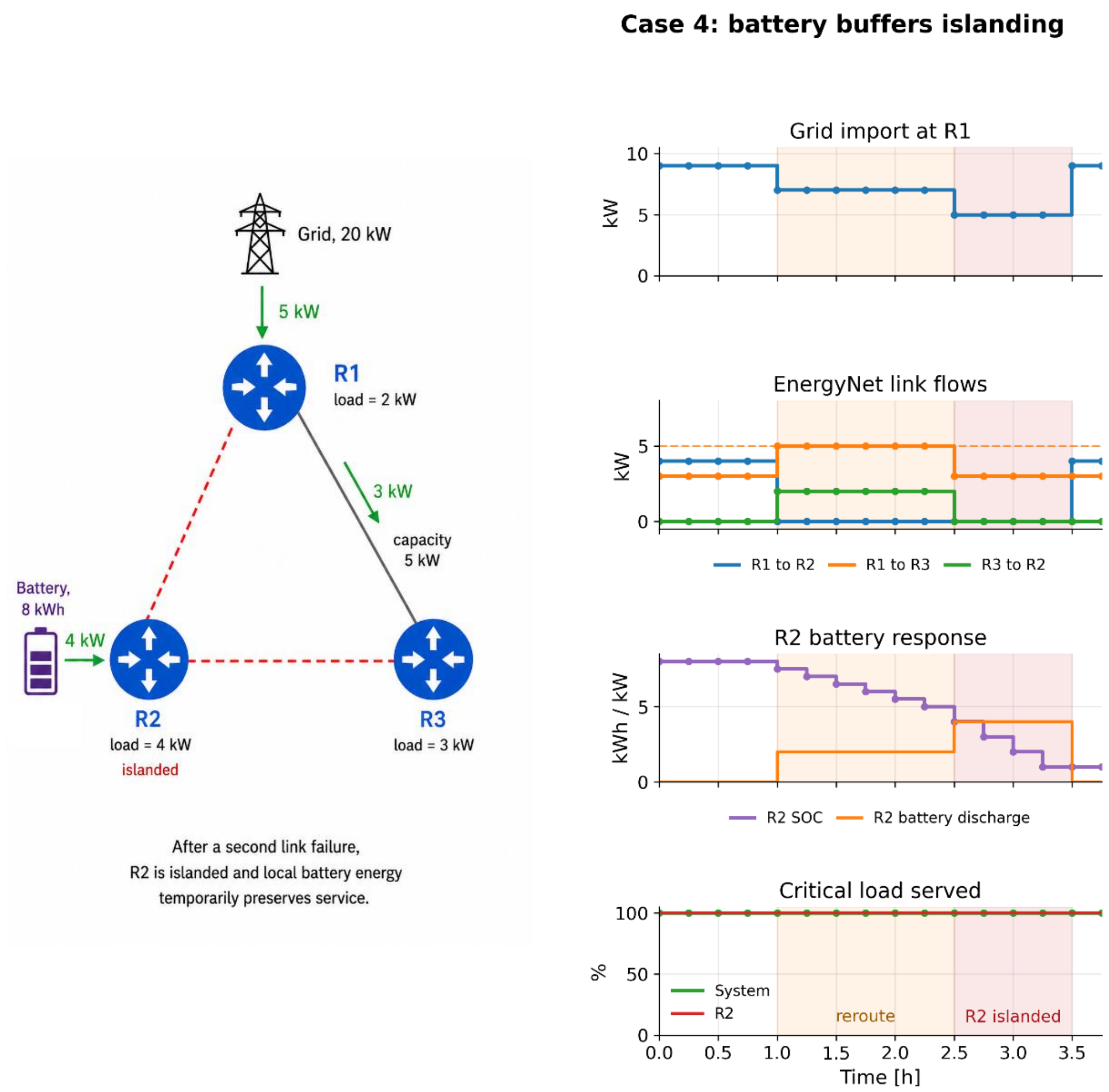

**Figure 11:** Operation with stationary energy storage at $R_2$ during capacity-limited rerouting and subsequent complete islanding. The storage system supplies the power that cannot be transferred through the constrained or unavailable Energy Links, maintaining uninterrupted critical-load service throughout the disturbance.

This case extends the previous example from graceful degradation to local resilience through storage. During constrained network operation, storage complements the power that can still be transferred through the EnergyNet; during complete islanding, it temporarily replaces network supply altogether. The example therefore illustrates how network redundancy and local energy storage provide complementary mechanisms for maintaining service under progressively more severe disturbances.

### 5.6 Local generation and storage during islanding

The fifth case builds on the islanding condition introduced in Section 5.5 by adding local photovoltaic generation at $R_2$ as shown in Figure 12.

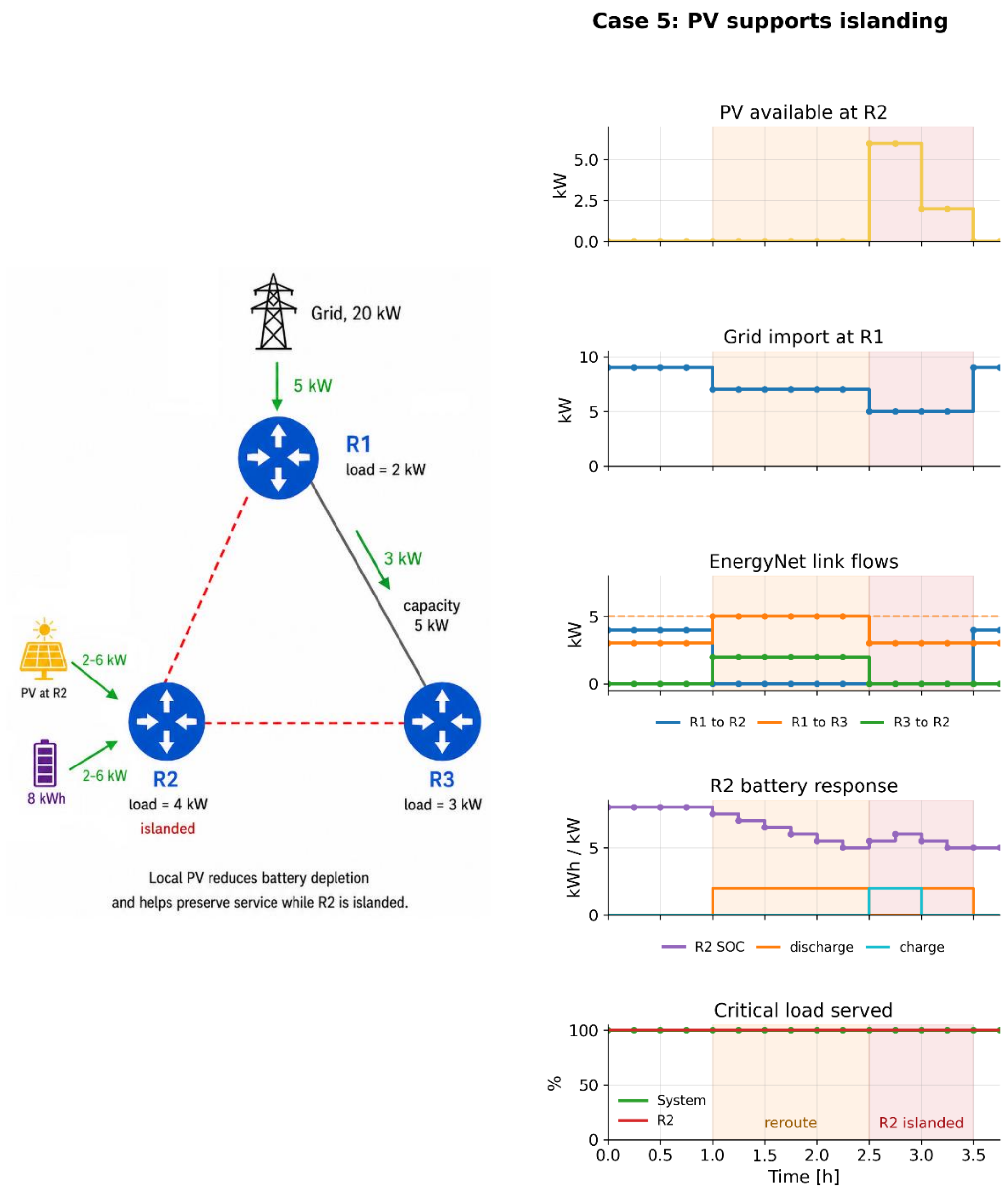

**Figure 12:** Operation with local generation and stationary storage at the islanded Energy Router $R_2$. The optimization coordinates local generation, consumption, and storage charging and discharging to satisfy the local critical load while preserving stored energy for later use.

During the initial capacity-constrained period, the EnergyNet, local generation, and storage jointly supply the critical load at $R_2$. When $R_2$ becomes completely islanded at $t = 2.50$h, its critical demand must instead be supplied entirely from local resources. Available photovoltaic generation supplies the load directly, while the battery compensates for any remaining imbalance between generation and demand.

The addition of local generation therefore reduces the energy that must be drawn from storage during the disturbance and extends the period for which the islanded router can sustain its load. More generally, the case illustrates the complementary roles of network connectivity, local generation, and stored energy: the network provides access to resources elsewhere in the EnergyNet, generation provides energy locally when available, and storage bridges temporal mismatches between generation and demand.

### 5.7 Sharing surplus local generation

The final case returns to the normal network topology, with all Energy Links available at their original 20 kW capacity and no storage or outages. Local generation of 12 kW is introduced at $R_2$ between $t = 1.00$ h and $t = 2.50$ h, exceeding the total 9 kW demand of the system, Figure 13.

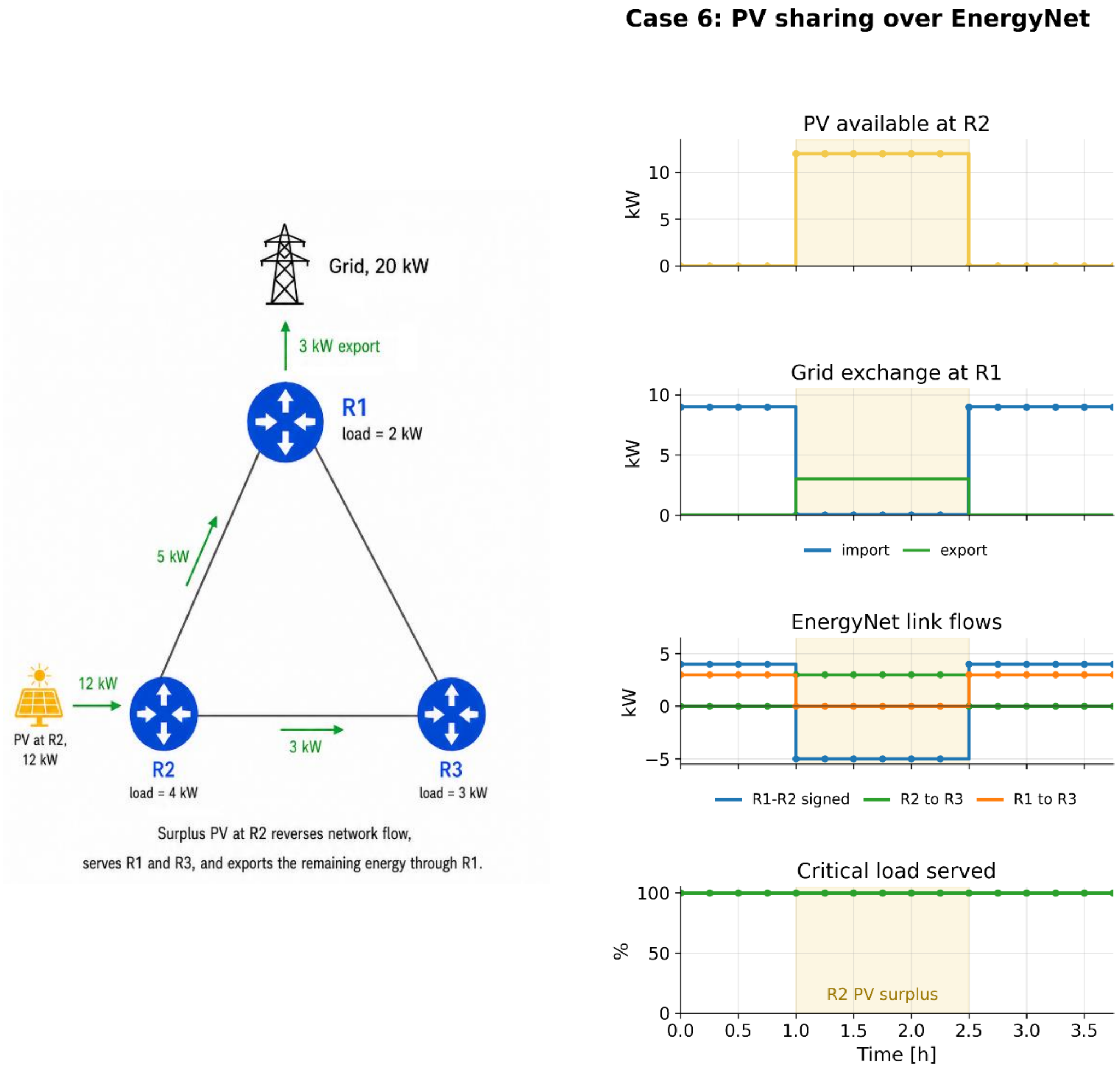

**Figure 13:** Operation with surplus local generation at Energy Router $R_2$. The local generation supplies demand at $R_2$, while the remaining energy is transferred through the EnergyNet to supply demand at $R_1$ and $R_3$ and export the remaining surplus through the public-grid interface at $R_1$.

Before local generation becomes available, operation is identical to the reference case, with $R_1$ importing power from the public grid to supply the network. When generation at $R_2$ increases to 12 kW, 4 kW supplies its local load, while the remaining 8 kW is shared through the EnergyNet. Of this surplus, 3 kW supplies $R_3$, 2 kW supplies $R_1$, and the remaining 3 kW is exported through the public-grid interface at $R_1$. Grid import therefore falls to zero during the generation interval. When local generation ceases, the original power-flow pattern is restored.

This final case illustrates how distributed generation can be utilized as a network-level resource rather than solely a local resource. Once the demand at $R_2$ is satisfied, its surplus generation supplies demand elsewhere in the EnergyNet, displacing public-grid import before the remaining surplus is exported. The EnergyNet thereby enables locally available generation to contribute to the energy balance of the wider network, allowing resources distributed across individual routers to be coordinated at system level.

### 5.8 Summary

The three-router cases illustrate how the same EnergyNet model and dispatch formulation respond as the available resources and network conditions change. Energy can be transferred between routers during normal operation, redirected through alternative network paths following link outages, and limited by the available transfer capacity when the network becomes constrained. Local storage can supplement insufficient network transfer and sustain loads during temporary islanding, while local generation can supply demand, preserve stored energy, and provide surplus energy to the wider network.

These mechanisms arise from the interaction of network connectivity, transfer capacity, generation, storage, grid access, and load demand within the same underlying formulation. No separate operating mode is prescribed for the individual scenarios; the dispatch is recalculated according to the resources, topology, constraints, and priorities present at each timestep.

## 6. Application: mobile energy peering between geographically distributed EnergyNets

The preceding case demonstrated how distributed EnergyNets can coordinate their available generation, storage, and loads to support one another. We now consider a practical realization of these principles based on two existing microgrid installations: a grid-connected system at Stanford University, [20]-[21], and an associated off-grid system located at Half Moon Bay south of San Francisco, Figure 14. Both sites are independently functioning energy systems with local generation and storage. In the considered EnergyNet configuration, the Stanford Farm and Half Moon Bay microgrids are coordinated together with an electric transit bus operating from Stanford [22]-[24], Figure 14. The bus retains its primary transportation function, while its onboard storage and bidirectional charging capability enable it to provide scheduled energy support to Half Moon Bay when needed.

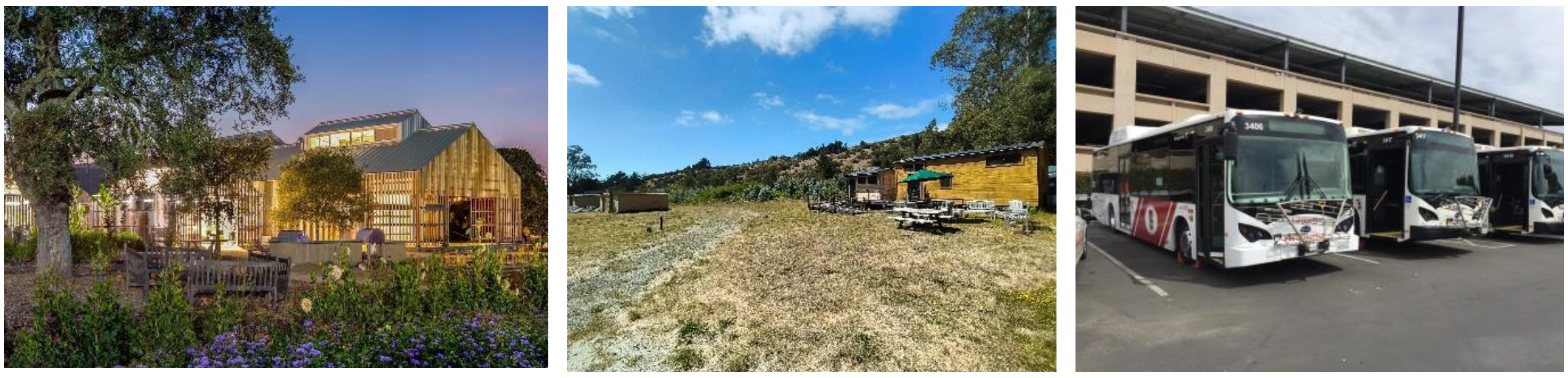

**Figure 14:** Existing microgrid installations and transportation infrastructure considered in the application case: the Stanford Farm, the off grid microgrid at Half Moon Bay, and the campus electric bus infrastructure

### 6.1 Configuration

The configuration considered in this study is summarized in Figure 15 and Table 2. The Stanford Farm microgrid comprises a 35 kW photovoltaic installation, 45 kWh of stationary battery storage, local loads, and a connection to the public electricity grid. The electric bus is part of the Stanford campus transit system and is normally served by the existing nearby charging depot. The Half Moon Bay microgrid operates off-grid and comprises approximately 20 kW of photovoltaic generation, 60 kWh of stationary battery storage, and local loads.

The electric bus provides the mobile energy resource within the considered configuration. Its battery has a nominal capacity of 350 kWh, of which a maximum of 315 kWh is made available during routine operation, corresponding to 90% of nominal capacity. A minimum mobility reserve of 70 kWh is maintained, leaving an operating energy window of 245 kWh. At Half Moon Bay, a bidirectional interface allows energy to be transferred from the vehicle to the microgrid during a scheduled support visit.

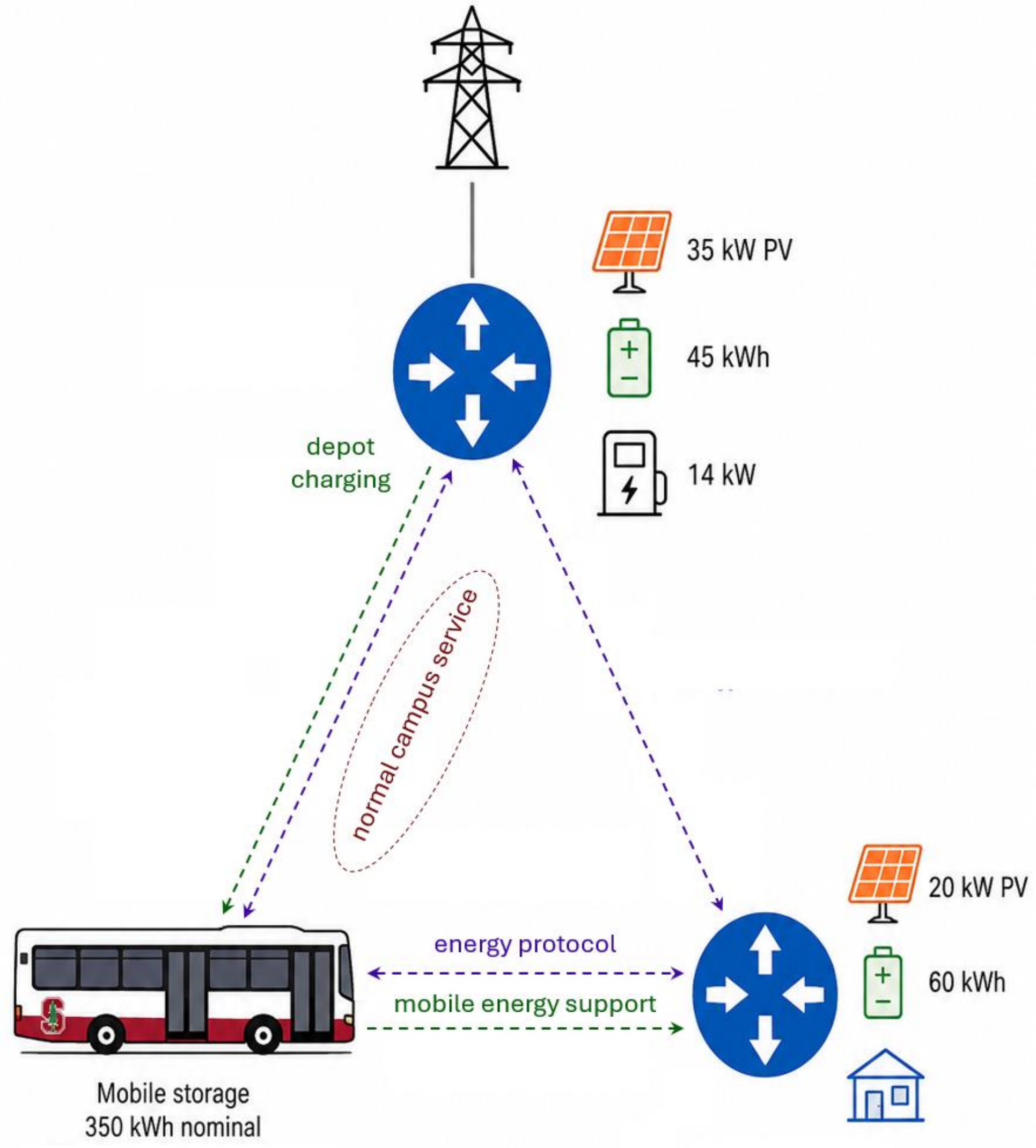

**Figure 15:** EnergyNet representation of the Stanford Farm - Half Moon Bay application. The two microgrids coordinate their local energy resources through the Energy Protocol, while the electric bus provides a mobile energy resource that enables scheduled physical energy exchange between the geographically separated sites.

**Table 2: System parameters adopted for the Stanford Farm - Half Moon Bay application**

| Component | The Farm | Half Moon Bay | Mobile resource |
|---|---|---|---|
| Grid connection | Yes | None | N/A |
| PV capacity | 35 kW | 20 kW | N/A |
| Stationary storage | 45 kWh | 60 kWh | N/A |
| Stationary battery power | 15 kW | 20 kW | N/A |
| Bidirectional interface | 2 x 11.5 kW | 50 kW | N/A |
| Vehicle capacity | - | - | 350 kWh |
| Mobility reserve | - | - | 70 kWh |
| Depot charging | - | - | 30 kW |
| Route-energy requirement | - | - | 150 kWh/day |

Within the EnergyNet framework, the two microgrids and the mobile storage resource are coordinated according to their available and anticipated energy states. In the scenario considered here, a period of low solar generation is forecast at Half Moon Bay, indicating that its stationary storage will progressively decline. Through the Energy Protocol, the Half Moon Bay microgrid communicates its anticipated energy requirement, allowing a support visit by the electric bus to be scheduled before the local storage becomes depleted.

The resulting operational sequence is evaluated over a three-day simulation with a temporal resolution of 30 min. During each day, the bus first performs its normal campus transportation duty before traveling from Stanford to Half Moon Bay. It connects to the Half Moon Bay microgrid for approximately one hour, during which energy can be transferred through the bidirectional interface, and subsequently returns to Stanford. The vehicle then begins continuous overnight charging period at the campus depot before returning to transportation service the following day. This sequence is repeated once per day throughout the simulation.

### 6.2 Simulation results

The resulting three-day operation is shown in Figure 16. Under the assumed low-solar conditions, corresponding to 20% of the clear-day photovoltaic profile, the Half Moon Bay microgrid battery progressively discharges as local generation becomes insufficient to fully replenish the energy consumed by the load. Based on the forecasted evolution of the battery state, a recovery target is communicated, and a preventive support visit is scheduled for each day.

The bus continues its normal transportation operation before traveling to Half Moon Bay and connecting to the local microgrid. Across the three scheduled visits, a total of around 21 kWh is transferred to Half Moon Bay. Following each visit, the bus returns to Stanford and begins its overnight depot charging cycle. Despite its combined transportation and energy-support duties, the vehicle reaches a minimum stored energy of around 157 kWh, remaining well above the imposed 70 kWh mobility reserve.

The effect of the preventive intervention is evident in the Half Moon Bay storage trajectory. Without mobile support, the stationary battery declines to near-zero by the end of the simulation. With the scheduled support visits, it instead reaches the communicated recovery level; the mobile support prevents the progressive erosion of the remaining energy reserve while preserving the primary transportation function of the vehicle.

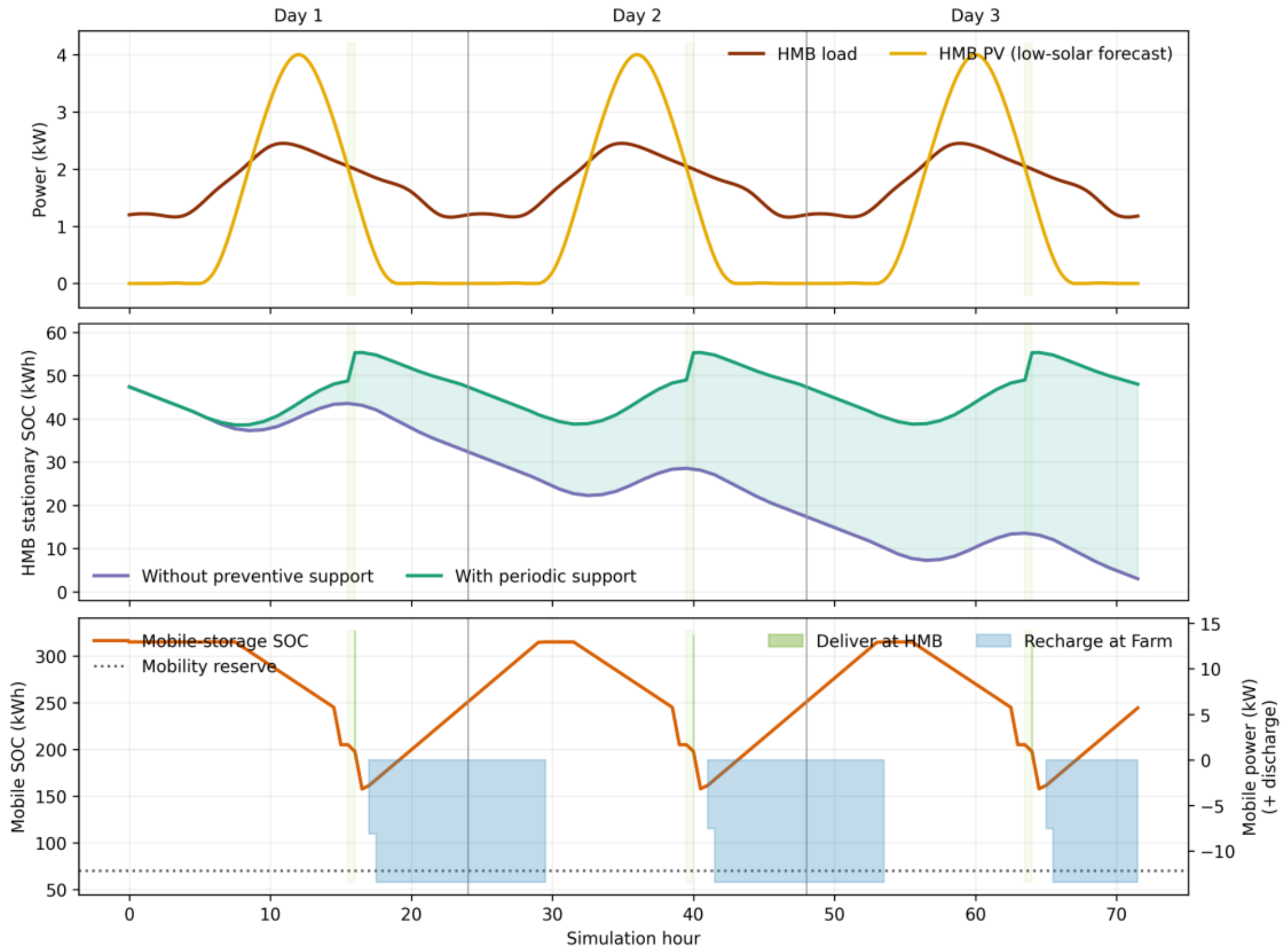

**Figure 16:** Simulated three-day operation of the Stanford Farm - Half Moon Bay EnergyNet configuration under low solar conditions. The results show the scheduled bus operation and energy-support visits, vehicle battery state and overnight depot charging, and the Half Moon Bay stationary battery with and without preventive mobile support.

### 6.3 Discussion

The results demonstrate a practical application of the EnergyNet coordination principles established in the preceding cases. The Stanford Farm and Half Moon Bay systems remain independently functioning microgrids, while coordination of the electric bus provides additional energy flexibility when required. Importantly, this support does not replace the primary transportation function of the vehicle. The bus completes its normal operating duty, provides a limited period of bidirectional energy support, and is subsequently replenished through the existing depot charging infrastructure.

The case also illustrates the value of anticipatory coordination. Rather than responding to a loss of supply, the forecasted decline in Half Moon Bay storage is identified in advance and the available mobile resource is scheduled accordingly. Repeated short support periods are sufficient to maintain the requested storage level while preserving the vehicle mobility reserve. Bidirectional charging therefore allows electric vehicles to participate as time-dependent EnergyNet resources without subordinating their primary mobility function. Their value as resilience resources is determined not only by available battery capacity, but also by where and when they can connect to the EnergyNet.

Whereas this case considers intermittent energy exchange through a mobile resource, the following section examines a permanently interconnected, neighbourhood-scale EnergyNet comprising ten Energy Routers and multiple fixed paths for energy exchange.

## 7. Application: quantifying resilience of a neighborhood EnergyNet

The framework developed in the preceding sections is now applied to an under-construction neighborhood-scale EnergyNet of ten interconnected apartment buildings comprising 278 apartments. The study evaluates the system under realistic time-varying demand and photovoltaic generation, distributed battery storage, multiple grid interfaces, and network disturbances.

The analysis progresses from normal representative summer and winter operation to increasingly severe disturbances involving loss of grid and Energy Link infrastructure and partial network islanding. Long-horizon reliability screening is then used to characterize component-failure exposure, followed by Monte Carlo consequence analysis to quantify the service maintained during a severe disturbance under uncertain operating conditions.

Throughout this section, resilience is evaluated primarily through the ability of the EnergyNet to maintain critical electrical service during disturbances. The corresponding system response is examined through the redistribution and use of grid supply, local generation, storage, and Energy Link transfers as network capability is progressively reduced.

### 7.1 Configuration

The investigated system is an under-construction commercially specified neighborhood-scale EnergyNet in Lund, Sweden, with ten apartment buildings comprising 278 apartments, as shown in Figure 17. The buildings are represented by ten Energy Routers $R_1, \dots, R_{10}$, interconnected by ten Energy Links in a closed ring topology. Public-grid interfaces are located at $R_1$ and $R_6$, providing two external connection points on opposite sides of the ring. Each router is additionally equipped with local photovoltaic generation and battery storage. The resulting EnergyNet configuration is shown in Figure 18, and the principal system parameters are summarized in Table 3.

The aggregate residential demand is specified in the range of 900–1800 kWh/day, with an additional service demand of approximately 200 kWh/day. The seasonal demand profiles used in the simulations are scaled to an annual demand of 650 MWh, of which approximately 75% is classified as critical and 25% as flexible. Photovoltaic generation is based on PVGIS solar data, [25], and a total installed capacity of 500 kWp, distributed equally among the ten routers.

Each router has 100 kWh of battery storage with a 50 kW power rating and an operating SOC range of 10–90%. Each Energy Link is rated at 100 kW, while the two public-grid interfaces have a combined maximum capacity of 600 kW for import and 300 kW for export. The deterministic simulations use a 30 minute timestep and a 24 hour planning horizon. Demand and photovoltaic generation are provided as forecasts over this horizon, while component failures are not anticipated and enter the dispatch only when they occur.

The system is evaluated as specified; optimization of component sizing and network topology is outside the scope of the present study.

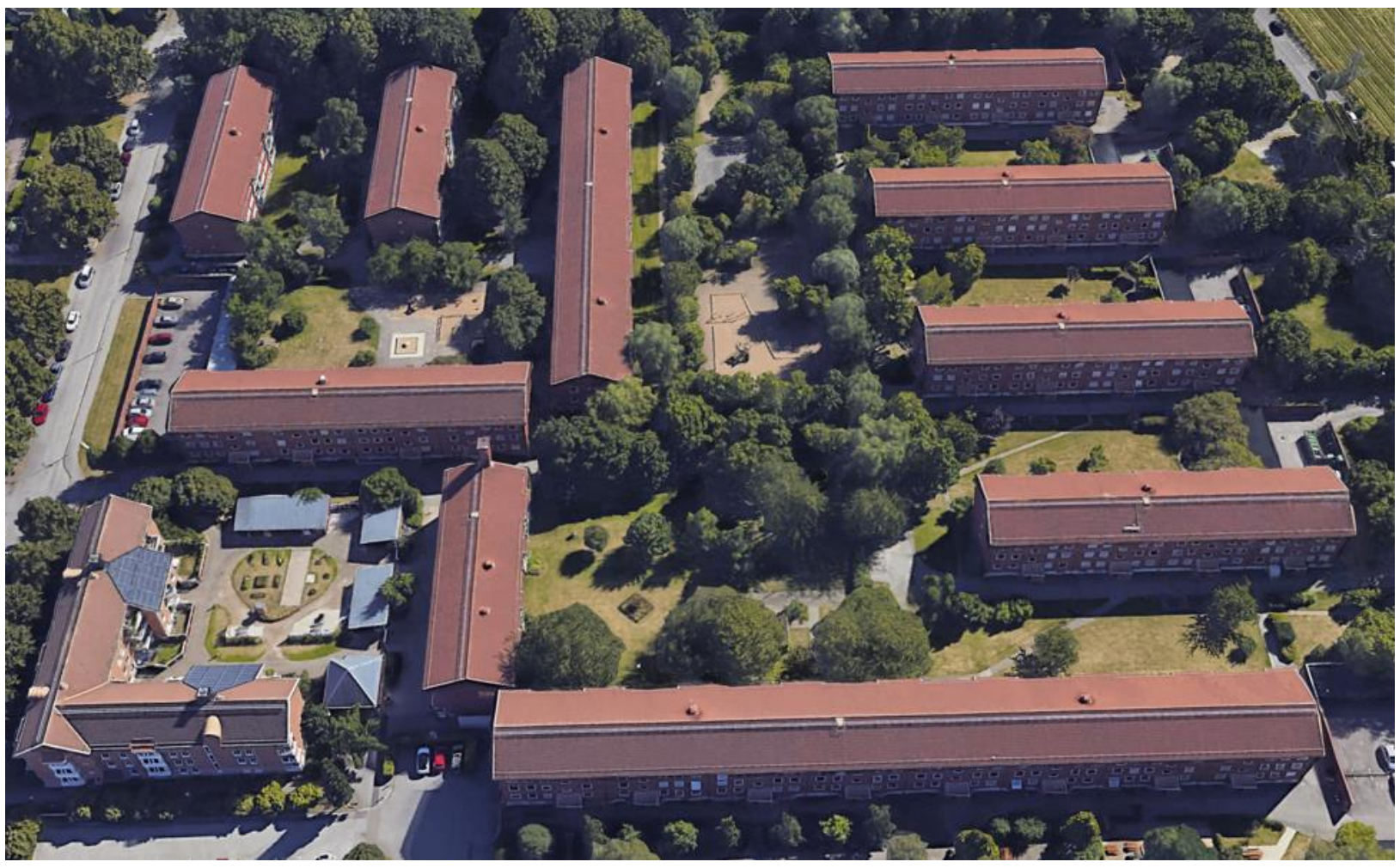

**Figure 17:** Overview of commercial proof-of-concept EnergyNet installation in Lund, Sweden.

**Table 3:** Configuration of the simulated neighborhood-scale EnergyNet

| **Parameter** | **Value** |
|---|---|
| Buildings / Energy Routers | 10 |
| Apartments | 278 |
| Annual electrical demand | 650,000 kWh |
| Energy Link rating | 100 kW each |
| Public-grid interfaces | $R_1, R_6$ |
| Maximum grid import | 300 kW per interface |
| Maximum grid export | 150 kW per interface |
| Photovoltaic capacity | 50 kWp per router (500 kWp total) |
| Battery capacity | 100 kWh per router (1,000 kWh total) |
| Battery power | 50 kW per router (500 kW total) |
| Battery SOC operating range | 10-90% |
| Simulation timestep | 30 minutes |
| Planning horizon | 24 hours |
| Critical demand | ~75% of annual demand |
| Flexible demand | ~25% of annual demand |
| Simulation period | 72 hours |

### 7.2 Normal operation

Before introducing disturbances, the EnergyNet is evaluated during representative 72-hour summer and winter periods. These cases establish the normal operating behavior of the system under contrasting levels of photovoltaic generation and provide the reference conditions for the subsequent disturbance studies.

*Summer high-solar operation*

Figure 18 shows the EnergyNet during a representative high-solar summer period. All critical and flexible demand is supplied throughout the simulation. Photovoltaic generation exceeds total electrical demand over the 72-hour period and, together with the distributed storage, eliminates grid import throughout the simulation.

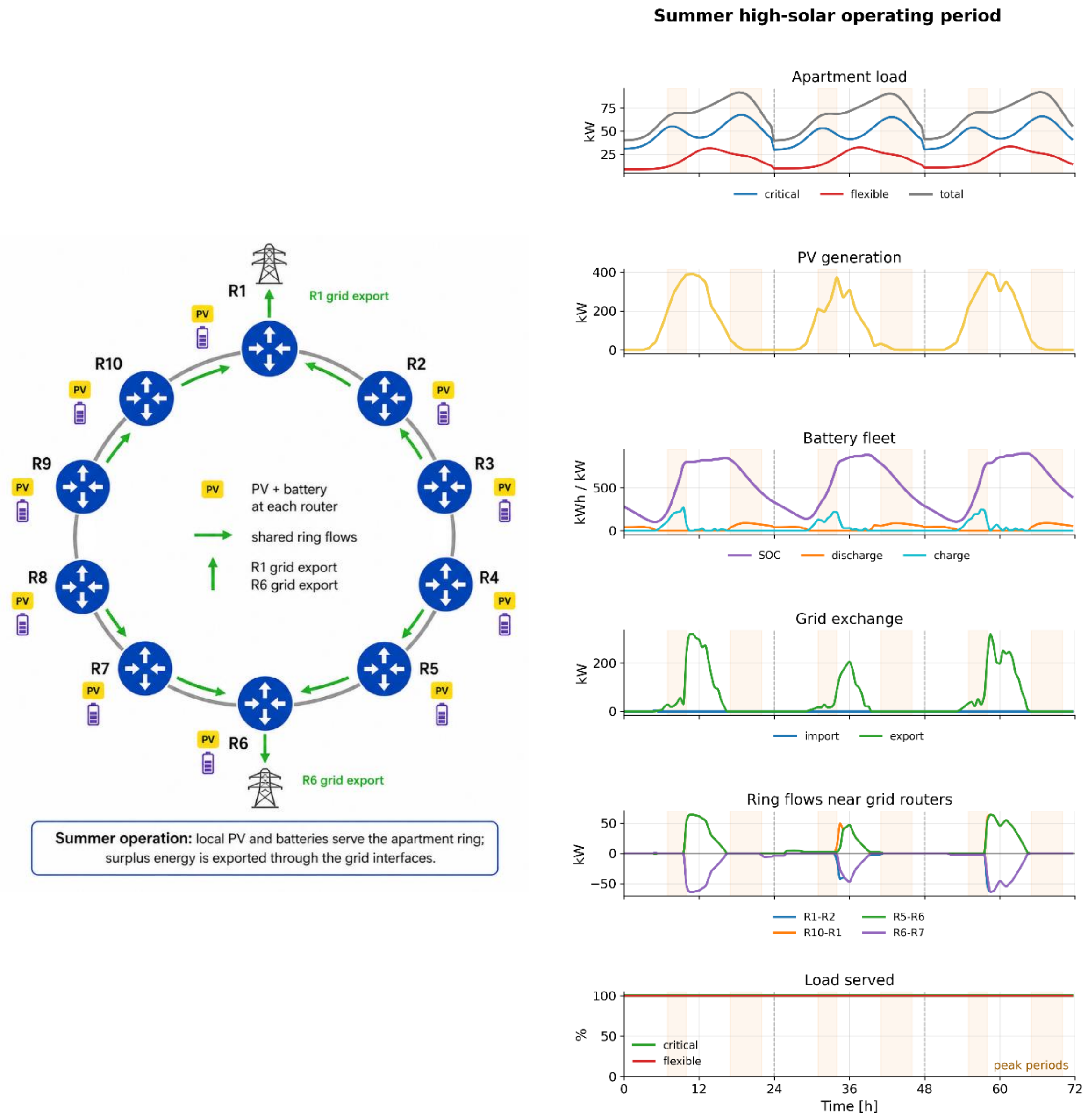

**Figure 18:** Normal summer operation of the commercial EnergyNet. Representative 72 hour operating period showing electrical demand, photovoltaic generation, battery state of charge, public-grid exchange, and Energy Link utilization under high solar generation.

During periods of high photovoltaic generation, local generation supplies demand and charges the distributed batteries. Stored energy is subsequently discharged as photovoltaic generation decreases, shifting part of the daytime surplus to periods of lower generation. Surplus energy that cannot be utilized by local demand or storage is transferred through the EnergyNet and exported through the public-grid interfaces at $R_1$ and $R_6$.

The Energy Link flows reflect the spatial redistribution of generation, demand, storage, and grid exchange across the network. All transfers remain well within the installed ratings, with the highest Energy Link loading reaching approximately 64%. The summer case therefore establishes a generation-surplus operating regime in which distributed photovoltaic generation and storage supply the neighborhood demand while the remaining energy is exported to the public grid.

*Winter low-solar operation*

Figure 19 shows the EnergyNet during a representative 72-hour winter period with substantially lower photovoltaic generation. All critical and flexible demand remains supplied throughout the simulation, but the reduced local generation shifts the system from the generation-surplus operation observed in summer to a grid-supported energy-deficit regime.

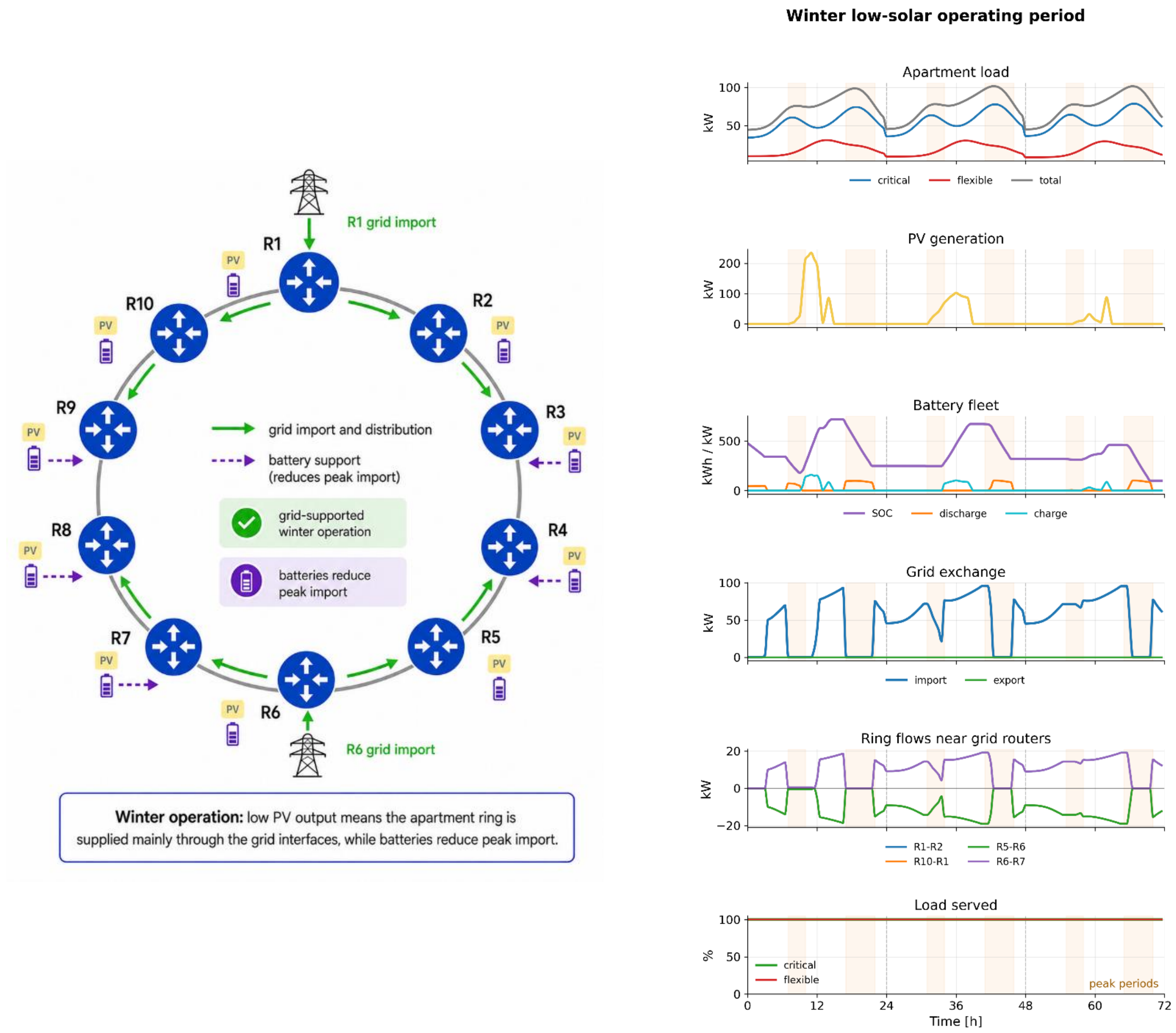

**Figure 19:** Normal winter operation of the commercial EnergyNet. Representative 72-hour operating period showing electrical demand, photovoltaic generation, battery state of charge, public-grid exchange, and Energy Link utilization under low solar generation.

Photovoltaic generation contributes when available, while the remaining demand is supplied primarily through the public-grid interfaces at $R_1$ and $R_6$. The distributed batteries charge and discharge over the 24 h planning horizon, shifting energy between periods, while Energy Link flows distribute power between routers according to their local energy balances.

Despite the greater dependence on external supply, the available grid and Energy Link capacities remain well above the required power transfers. The winter case therefore establishes a normal low-solar operating condition in which grid import, local generation, storage, and inter-router energy transfer maintain full electrical service without approaching the system transfer limits.

### 7.3 Disturbance scenarios

Having established the normal operating behavior of the EnergyNet, three deterministic disturbance scenarios are introduced to examine its response as network capability is progressively reduced. The cases consider the loss of one public-grid interface, the loss of one internal Energy Link, and simultaneous Energy Link failures resulting in partial network islanding. The progression therefore moves from reduced access to external supply, through reduced internal transfer capability, to complete loss of network access for part of the EnergyNet.

Each simulation covers 72 h, with the disturbance introduced at hour 24 and the affected components restored at hour 42, representing an 18-hour outage. This provides normal operating periods before and after each disturbance and allows the transition into, operation during, and recovery from the fault condition to be observed. The disturbances are not included in the forecasts before they occur: component availability changes only when the fault is realized. Following the fault, the assumed restoration time is available to the receding-horizon dispatch. The pre-fault operation therefore follows the corresponding intact system trajectory, while the subsequent dispatch adapts to the changed network configuration.

*Loss of one public-grid interface*

The first disturbance removes the public-grid interface at $R_1$, leaving the interface at $R_6$, all Energy Links, photovoltaic generation, and distributed battery storage available. The $R_1$ Energy Router itself remains operational and connected to its neighboring routers. The disturbance therefore reduces external grid access from two connection points to one without changing the internal topology of the EnergyNet.

Figure 20 shows that all critical and flexible demand remains supplied throughout the outage. At the onset of the disturbance, grid exchange at $R_1$ falls to zero and external supply shifts to $R_6$. The corresponding Energy Link flows redistribute as energy entering through the remaining grid interface is transferred across the network to supply the distributed loads. When the $R_1$ grid interface is restored, the system returns to operation with both external connection points available.

The disturbance therefore produces a substantial change in the spatial distribution of power without affecting electrical service. Before the fault, external energy can enter the EnergyNet through both $R_1$and $R_6$. During the outage, $R_6$ becomes the sole public-grid interface, and the Energy Link flows reorganize to distribute this energy across the complete network. All critical and flexible demand remains supplied throughout the 18-hour disturbance.

The result is fundamentally a consequence of the remaining power and transfer capability. The surviving grid interface provides the required external supply, while the connected EnergyNet provides the paths needed to distribute it throughout the neighborhood.

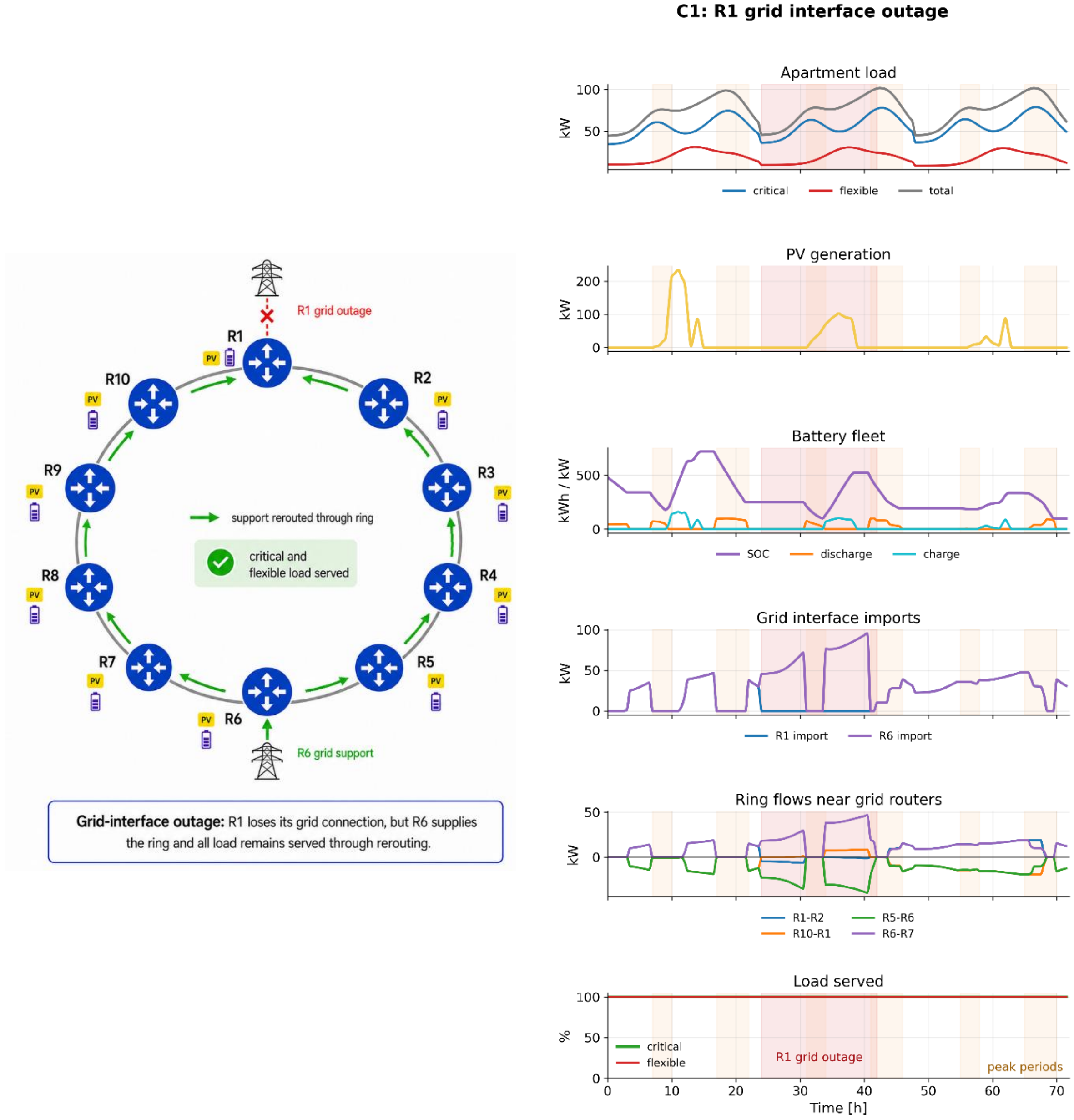

**Figure 20:** Loss of one public-grid interface. Operational response following the temporary loss of the public-grid connection at Router $R_1$, illustrating redistribution of power flows through the remaining grid interface while maintaining electrical service.

*Loss of one internal Energy Link*

The second disturbance removes the Energy Link between $R_5$ and $R_6$, while both public-grid interfaces and all other components remain available. Unlike the first case, external grid access is unchanged; instead, opening one link converts the closed ring into a connected path.

Figure 21 shows that all critical and flexible demand remains supplied throughout the outage. Power previously transferred across $R_5 - R_6$ is redirected around the surviving path through the opposite side of the ring. This redistribution is clearly visible in the Energy Link flows: $R_4 - R_5$ reverses direction, while links that previously carried little or no power take on the rerouted transfer. The largest Energy Link flow remains well below its rated capacity, and all loads remain fully supplied throughout the 18-hour disturbance.

The result shows the role of redundant transfer paths. Loss of a carrying Energy Link changes how energy moves through the EnergyNet but does not interrupt service while the remaining topology provides a connected path with sufficient transfer capacity.

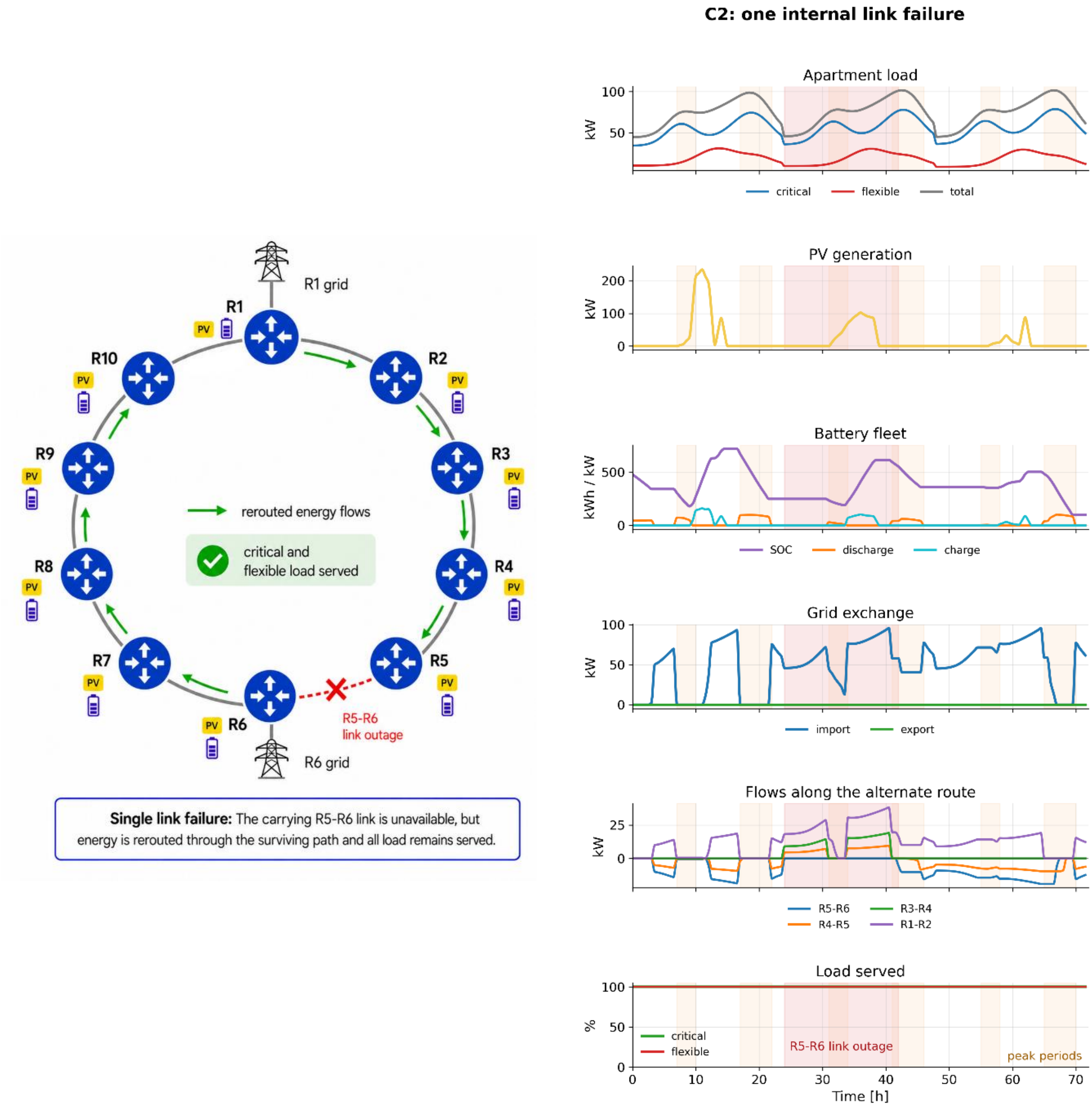

**Figure 21:** Loss of one Energy Link. Operational response following the temporary outage of the Energy Link between Routers $R_5$ and $R_6$, demonstrating automatic rerouting through the remaining network topology.

*Simultaneous loss of two internal Energy Links*

The third disturbance introduces simultaneous failures of the $R_3 - R_4$ and $R_5 - R_6$ Energy Links, isolating $R_4$ and $R_5$ from the remainder of the EnergyNet for 18 hours. Unlike the preceding cases, no surviving transfer path connects the affected routers to either public-grid interface. Boundary transfer is therefore zero throughout the disturbance, while the remaining eight routers retain full critical and flexible load service as shown in Figure 22.

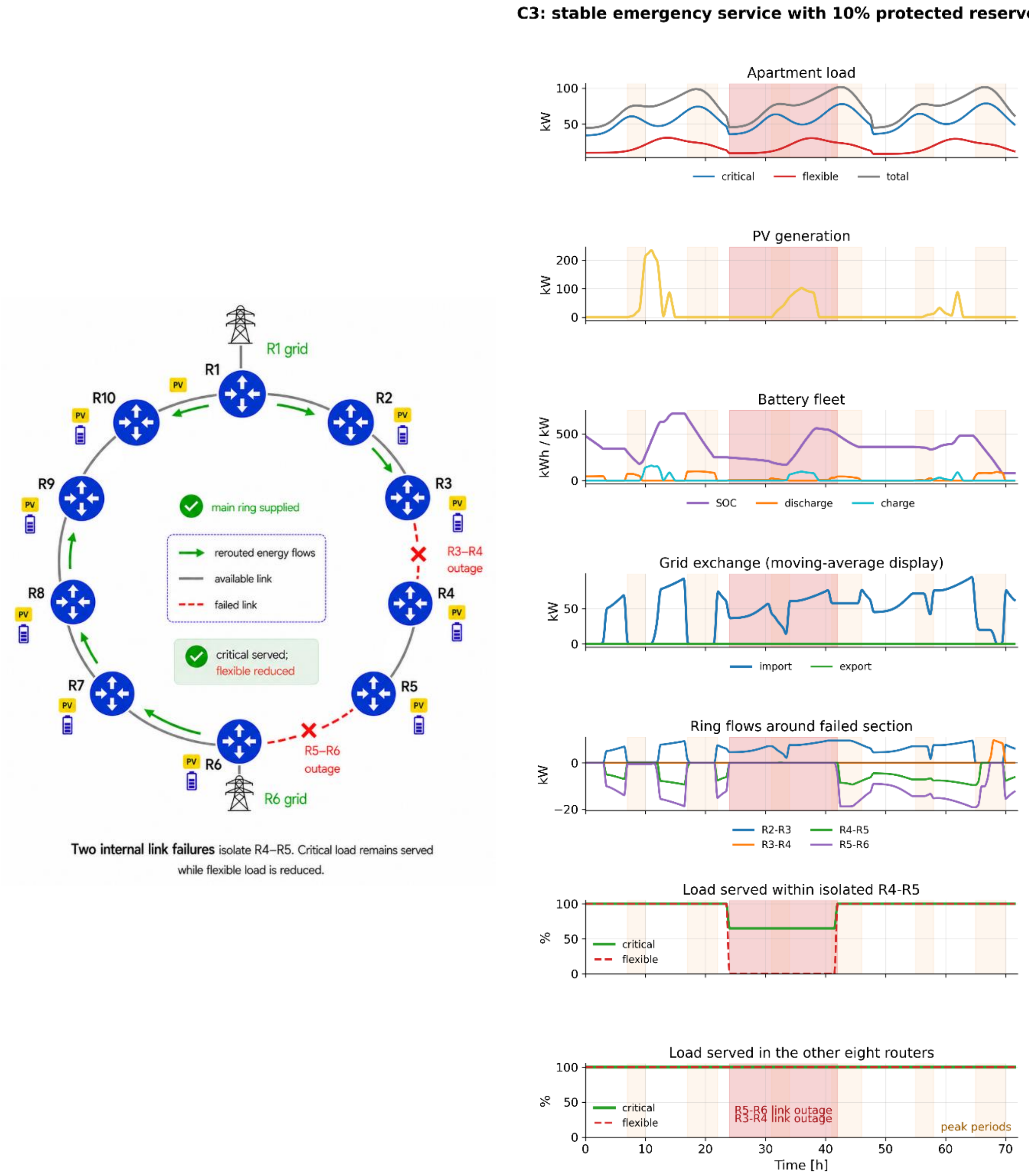

**Figure 22:** Simultaneous loss of two Energy Links. Operational response during temporary islanding of part of the EnergyNet following simultaneous outages of the $R_3$–$R_4$ and $R_5$–$R_6$ Energy Links, illustrating priority-based load allocation and battery-supported operation.

Within the island, electrical service must consequently be sustained entirely from local photovoltaic generation and stored energy until connectivity is restored. The emergency reserve and sustainable service commitment introduced in Section 3.4 are applied following the disturbance. For the case shown in Figure 22, a 10% battery reserve is protected before the fault. Based on the available energy and expected conditions over the repair interval, the island commits to maintaining 65% of its critical demand throughout the outage, while flexible demand is curtailed.

Figure 22 shows that this commitment is maintained throughout the 18-hour islanding period. The reduction in service is confined to $R_4$ and $R_5$; the remainder of the EnergyNet continues unaffected. The case therefore demonstrates controlled degradation when the topology successfully contains a disturbance, but the isolated region does not have sufficient local energy to maintain full service before restoration.

*Effect of protected emergency reserve*

The amount of stored energy protected before the disturbance directly affects the service that can be maintained after islanding. To quantify this relationship, the C3 disturbance is repeated over a range of protected battery reserve levels. Figure 23 reports both the total critical-load energy supplied during the outage and the critical-load fraction that can be maintained continuously throughout the complete repair interval.

The distinction between these measures is important. At low reserve levels, a greater fraction of critical energy can be supplied over the event than can be guaranteed continuously, because energy available later cannot compensate for service that could not be supplied earlier. With no protected reserve, around 70% of critical energy can be supplied in total, while only around 40% can be maintained continuously. This difference progressively closes as the reserve increases.

Full continuous critical service is reached at approximately 30% reserve. Beyond this point, additional stored energy becomes available to restore flexible demand. The required reserve depends on the operating conditions and disturbance duration and is therefore specific to the investigated case.

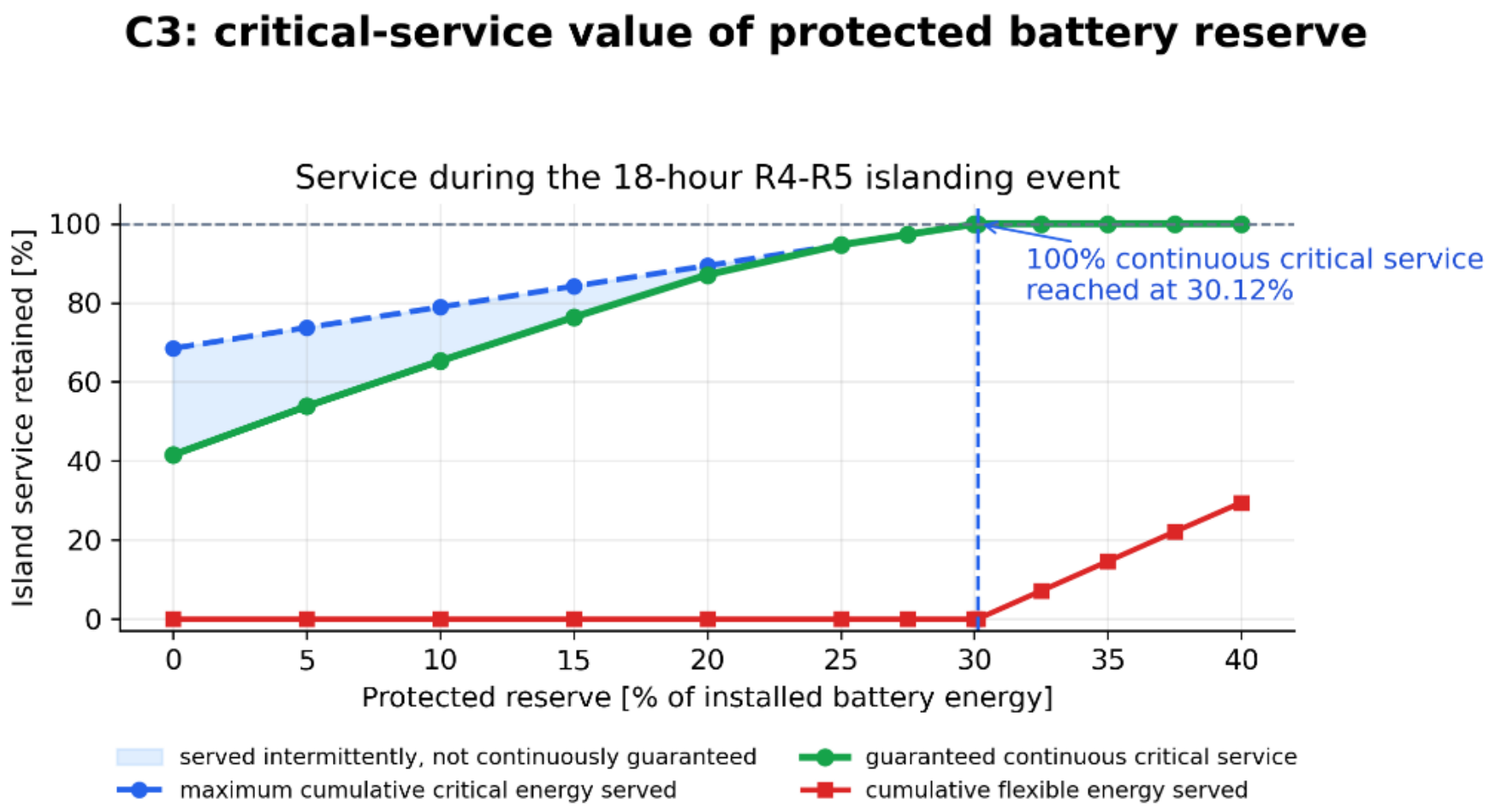

**Figure 23:** Effect of protected battery reserve on total and continuously maintained critical-load service during the C3 islanding event.

### 7.4 Long-horizon reliability screening and consequence analysis

The deterministic cases demonstrate how the EnergyNet responds to specific disturbances, but do not indicate how such fault conditions arise over the lifetime of the system. Long-horizon reliability screening is therefore used to characterize the occurrence and overlap of component outages and to identify fault conditions that warrant detailed consequence analysis following established methods for power system reliability assessment [19].

The following analysis considers 10,000 independent 30-year operating histories using assumed independent exponential failure models. Energy Link converters are assigned an MTBF of 100,000 hours and Energy Router converters an MTBF of 150,000 hours. Repair duration is generated using an exponential model with a nominal MTTR of 12 hours and bounded between 6 and 36 hours. These values are study assumptions used for reliability screening rather than field-derived reliability data.

Figure 24 summarizes the resulting long-horizon failure exposure. An average 30-year history contains approximately 44 individual component outages. Simultaneous component outages are considerably less frequent: approximately 8% of histories contain at least one overlapping failure, while about 3% contain overlapping Energy Link failures that partition the ring. A further 4.5% contain overlapping failures involving an Energy Router that increase the extent of network disconnection beyond that caused by either outage individually.

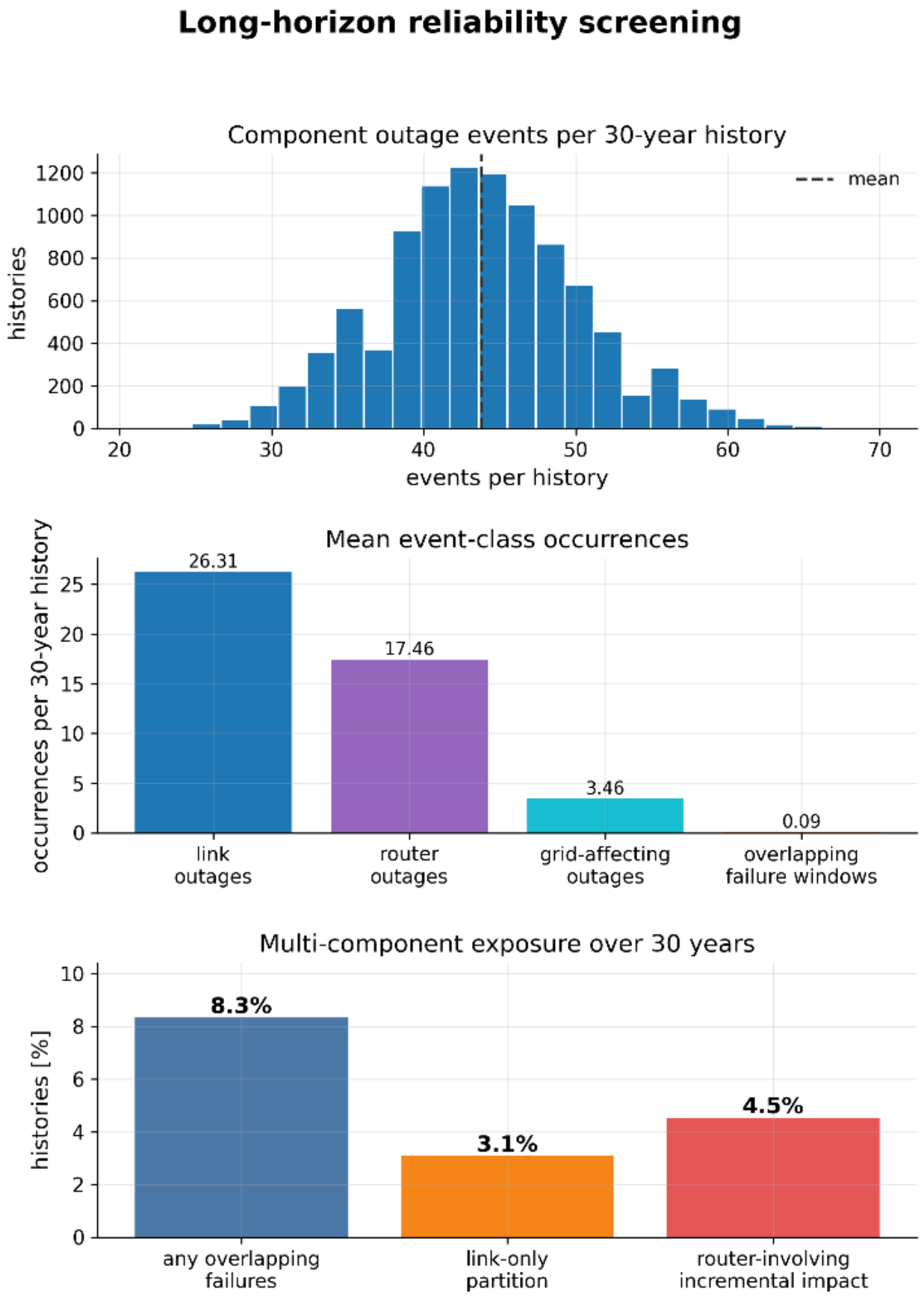

**Figure 24:** Long-horizon reliability screening results. Aggregated and event-class statistics obtained from 10,000 simulated 30-year operating histories, illustrating the occurrence and distribution of component outage events used to identify representative disturbance scenarios.

The screening therefore provides context for the consequence analysis. Network-partitioning events of the type considered in the preceding two-link failure case are relatively rare but remain credible over the lifetime of the system and represent a severe condition in which part of the EnergyNet loses access to both the public grid and the wider network. This disturbance is therefore used to examine how well critical service can be maintained when such an event occurs under varying operating conditions.

In the following, the two-link islanding event is retained while the conditions under which it occurs are varied across 1,000 realizations for both summer and winter operation. The sampled conditions include event timing, initial battery SOC, demand, photovoltaic generation, and repair duration. A 30% emergency reserve is protected before the disturbance, in accordance with Figure 23, and following islanding the emergency policy introduced in Section 3.4 determines the maximum critical-load fraction that can be maintained continuously over the repair interval.

Figure 25 shows the resulting critical-load service. Under summer conditions, 100% of critical demand is maintained in all 1,000 realizations. Winter conditions produce a broader range of outcomes as lower photovoltaic availability increases dependence on the energy stored within the island. The median continuously maintained critical-load service is approximately 92%, with around half of the realizations maintaining at least 90% of critical demand and more than 40% maintaining full critical service throughout the disturbance.

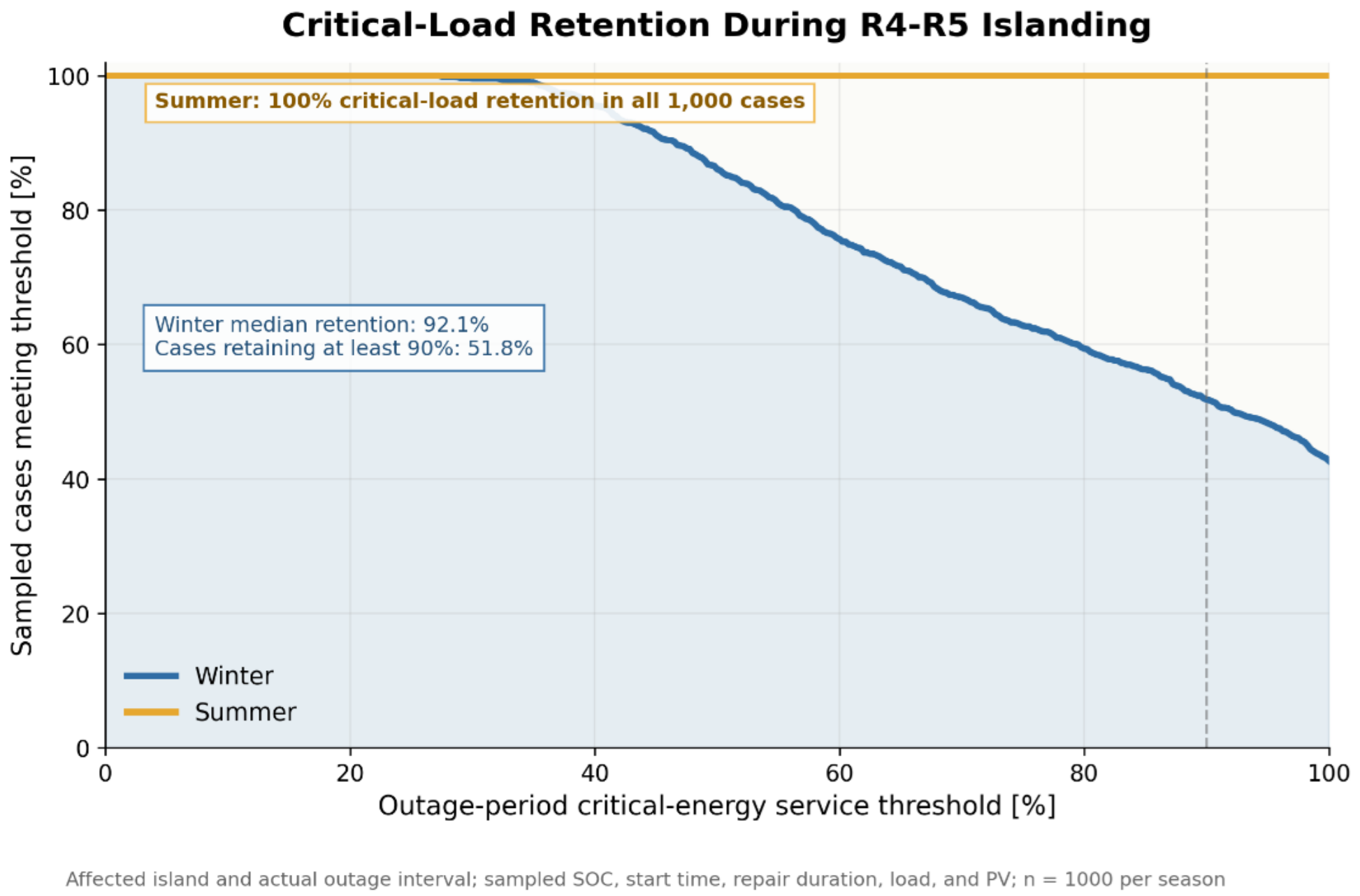

**Figure 25:** Monte Carlo consequence analysis of the two-link failure disturbance scenario. Distribution of critical-load retention obtained from 1,000 Monte Carlo realizations for representative summer and winter operating conditions, illustrating the probability of maintaining critical electrical service under uncertain operating conditions.

The results show that even for the severe network-partitioning event considered here, critical service can remain high across a wide range of operating conditions. Full service cannot always be maintained, particularly under unfavorable winter conditions, but degradation is controlled according to the available local energy and the duration of the disturbance. Together with the long-horizon screening, the analysis places these outcomes in context: the investigated partitioning event is relatively rare, and when it does occur, the EnergyNet can maintain substantial critical service even when complete service is not possible.

## 8. Conclusions

This work has presented a computational framework for the quantitative analysis of EnergyNet systems, combining a graph-based system representation with optimization-based dispatch, time-domain simulation, and stochastic analysis. The framework represents distributed demand, generation, storage, grid interfaces, and Energy Links within a common mathematical formulation, allowing physically feasible energy allocation and exchange to be evaluated across changing operating conditions and network configurations.

The illustrative three-router cases established the principal EnergyNet mechanisms in a transparent setting, including energy sharing, rerouting following link outages, graceful degradation under transfer constraints, and the complementary roles of network connectivity, local generation, and storage. The same framework was then applied to an under-construction commercially specified ten-router neighborhood EnergyNet under representative summer and winter operation and progressively more severe disturbances.

The neighborhood study shows that resilience depends on the combined adequacy of power, energy, and network transfer. Loss of one grid interface or one Energy Link was accommodated by the remaining supply and transfer capability without interrupting electrical service. Simultaneous loss of two Energy Links instead isolated two routers from the grid-connected remainder. The topology contained the disturbance, leaving the other eight routers fully supplied, while service within the island became dependent on local photovoltaic generation, stored energy, load priority, and repair duration.

The islanding case also demonstrates the role of operational policy. With a 10% protected battery reserve, the emergency policy maintained a constant 65% of critical demand throughout the 18-hour disturbance while flexible island demand was curtailed. The reserve analysis further distinguished between the total critical energy that can be supplied during an outage and the critical service that can be guaranteed continuously. For the investigated deterministic conditions, full continuous critical service was reached at approximately 30% protected reserve, after which additional stored energy increasingly supported flexible demand.

Long-horizon reliability screening placed the deterministic disturbances in context by identifying the component-outage combinations that arise under the assumed failure and repair models. Separate Monte Carlo consequence analysis retained the two-link islanding mechanism while varying its timing, repair duration, battery state, demand, and photovoltaic generation. Full critical service was maintained in all summer realizations, while winter performance was more variable but retained a median continuous critical service level above 90%.

The framework is intentionally general with respect to topology, component sizing, resource composition, and operational objective. It therefore provides a foundation for evaluating specific EnergyNet installations and for future studies of topology, storage and generation sizing, emergency policies, reliability, economics, and other decentralized energy applications.